\documentclass[twocolumn,superscriptaddress,showpacs,preprintnumbers,amsmath,amssymb]{revtex4}
\usepackage{amsmath}
\usepackage{amssymb}
\usepackage{dcolumn}
\usepackage{graphicx}
\usepackage{bm}
\usepackage{epstopdf}
\usepackage{threeparttable}
\usepackage{float}
\usepackage{color}
\usepackage{cleveref}

\usepackage{appendix}

\usepackage [english]{babel}
\usepackage [autostyle, english = american]{csquotes}
\MakeOuterQuote{"}
\usepackage[normalem]{ulem}

\makeatletter
\newcommand*{\rom}[1]{\expandafter\@slowromancap\romannumeral #1@}
\makeatother
\makeatletter
\def\p@subsection{}
\makeatother

\begin{document}
\title{Shear response of granular packings compressed above jamming onset}

\author{Philip Wang}
\affiliation{Department of Mechanical Engineering and Materials Science, Yale University, New Haven, Connecticut, 06520, USA}
\author{Shiyun Zhang}
\affiliation{Department of Mechanical Engineering and Materials Science, Yale University, New Haven, Connecticut, 06520, USA}
\affiliation{Department of Physics, University of Science and Technology of China, Hefei, Anhui 230026, China}
\author{Philip Tuckman}
\affiliation{Department of Physics, Yale University, New Haven, Connecticut, 06520, USA}
\author{Nicholas T. Ouellette}
\affiliation{Department of Civil and Environmental Engineering, Stanford University, Stanford, CA 94305, USA}
\author{Mark D. Shattuck}
\affiliation{Department of Physics and Benjamin Levich Institute, The City College of the City University of New York, New York, New York, 10031, USA}
\affiliation{Department of Mechanical Engineering and Materials Science, Yale University, New Haven, Connecticut, 06520, USA}
\author{Corey S. O'Hern}
\affiliation{Department of Mechanical Engineering and Materials Science, Yale University, New Haven, Connecticut, 06520, USA}
\affiliation{Department of Applied Physics, Yale University, New Haven, Connecticut, 06520, USA}
\affiliation{Department of Physics, Yale University, New Haven, Connecticut, 06520, USA}

\date{\today}

\begin{abstract}
We investigate the mechanical response of jammed packings of repulsive, frictionless spherical particles undergoing isotropic compression. Prior simulations of the soft-particle model, where the repulsive interactions scale as a power-law in the interparticle overlap with exponent $\alpha$, have found that the ensemble-averaged shear modulus $\langle G \rangle$ increases with pressure $P$ as $\sim P^{(\alpha-3/2)/(\alpha-1)}$ at large pressures. However, a deep theoretical understanding of this scaling behavior is lacking. We show that the shear modulus of jammed packings of frictionless, spherical particles has two key contributions: 1) continuous variations as a function of pressure along geometrical families, for which the interparticle contact network does not change, and 2) discontinuous jumps during compression that arise from changes in the contact network.  We show that the shear modulus of the first geometrical family for jammed packings can be collapsed onto a master curve: $G^{(1)}/G_0 = (P/P_0)^{(\alpha-2)/(\alpha-1)} - P/P_0$, where $P_0 \sim N^{-2(\alpha-1)}$ is a characteristic pressure that separates the two power-law scaling regions and $G_0 \sim N^{-2(\alpha-3/2)}$.  Deviations from this form can occur when there is significant non-affine particle motion near changes in the contact network.  We further show that $\langle G (P)\rangle$ is not simply a sum of two power-laws, but $\langle G \rangle \sim (P/P_c)^a$, where $a \approx (\alpha -2)/(\alpha-1)$ in the $P \rightarrow 0$ limit and $\langle G \rangle \sim (P/P_c)^b$, where $b \gtrsim (\alpha -3/2)/(\alpha-1)$ above a characteristic pressure $P_c$. In addition, the magnitudes of both contributions to $\langle G\rangle$ from geometrical families and changes in the contact network remain comparable in the large-system limit for $P >P_c$.

\end{abstract}

\maketitle

\section{Introduction} 
\label{intro}

Granular materials, such as collections of grains, bubbles, or other macroscopic particles, interact via highly dissipative forces, which cause these materials to come to rest unless they are continuously driven, e.g. by gravity, shear, or other applied deformations~\cite{granular}.  Further, granular materials transition from fluid- to solid-like states with a nonzero shear modulus when they are compressed to sufficiently large packing fractions~\cite{jamming}. Despite numerous experimental~\cite{jb}, theoretical~\cite{scaling}, and simulation studies~\cite{ciammarra} of the jamming transition in granular media, there are numerous open questions concerning the structural properties and mechanical response of jammed granular packings.

\begin{figure*}
    \centering
    \includegraphics[height=5.5cm]{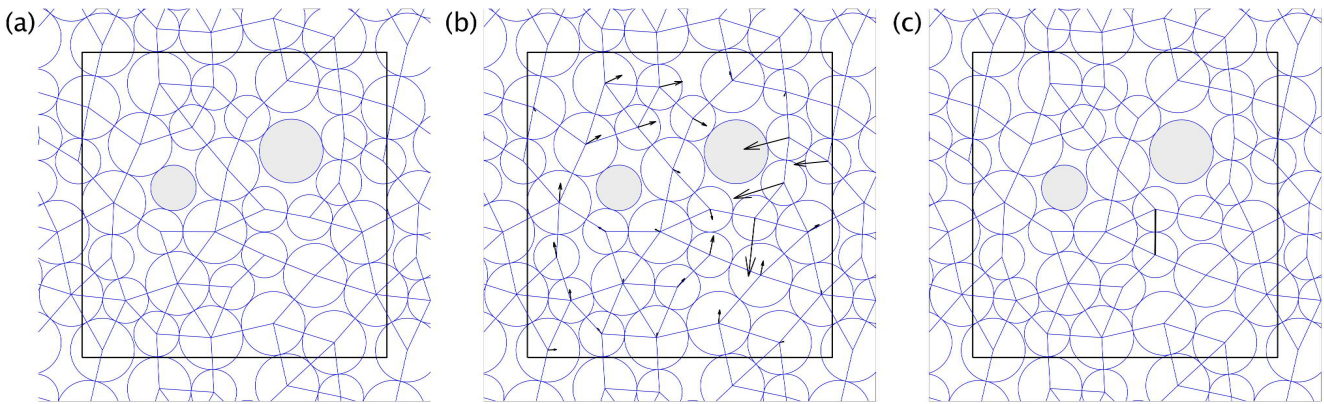}
    \caption{Snapshots of $N=32$ jammed disk packings with repulsive linear spring interactions ($\alpha=2$ in Eq.~\ref{potential}) undergoing isotropic compression before and after a point change where a new contact is added to the contact network (indicated by thin blue lines). In (a), we show an isostatic packing at $P=10^{-7}$ with $N_c=N_c^0 = 59$. (b) The packing in (a) has been compressed to $P=4.60 \times 10^{-6}$ without any changes in the contact network. The vectors indicate the displacements of the particle centers relative to the packing in (a) after multiplying by $100$. (c) The packing in (c) has been compressed to $P=4.65 \times 10^{-6}$, which results in the addition of one contact, $N_c = N_c^0+1$, indicated by the thick black line. The shaded particles indicate rattlers.}
    \label{fig:contactnetworkChange}
\end{figure*}

A simple, yet highly descriptive model for jamming in granular materials is one where we consider smooth, spherical particles that interact via the pairwise, purely repulsive, finite-ranged (``soft particle") potential~\cite{durian,makse}:
\begin{equation}
    \label{potential}
    U(r_{ij}) = \frac{\epsilon}{\alpha} \left( 1- \frac{r_{ij}}{\sigma_{ij}} \right)^{\alpha}
    \Theta \left( 1- \frac{r_{ij}}{\sigma_{ij}} \right),
\end{equation}
where $r_{ij}$ is the separation between the centers of particles $i$ and $j$, $\sigma_{ij} = (\sigma_i + \sigma_j)/2$ is the average diameter of particles $i$ and $j$, $\Theta(\cdot)$ is the Heaviside step function that prevents particles from interacting if they are not in contact,
$\epsilon$ is the characteristic energy scale, and $\alpha$ is the power-law scaling exponent of the interaction. For this interaction potential, the onset of jamming occurs when the number of interparticle contacts reaches the isostatic value~\cite{witten}, $N_c = N_c^0 = dN -d +1$, and the total pressure of the system $P > 0$, where $d=2$, $3$ is the spatial dimension and $N$ is the number of non-rattler particles~\cite{rattler}.

\begin{figure}[h]
    \centering
    \includegraphics[height=12cm]{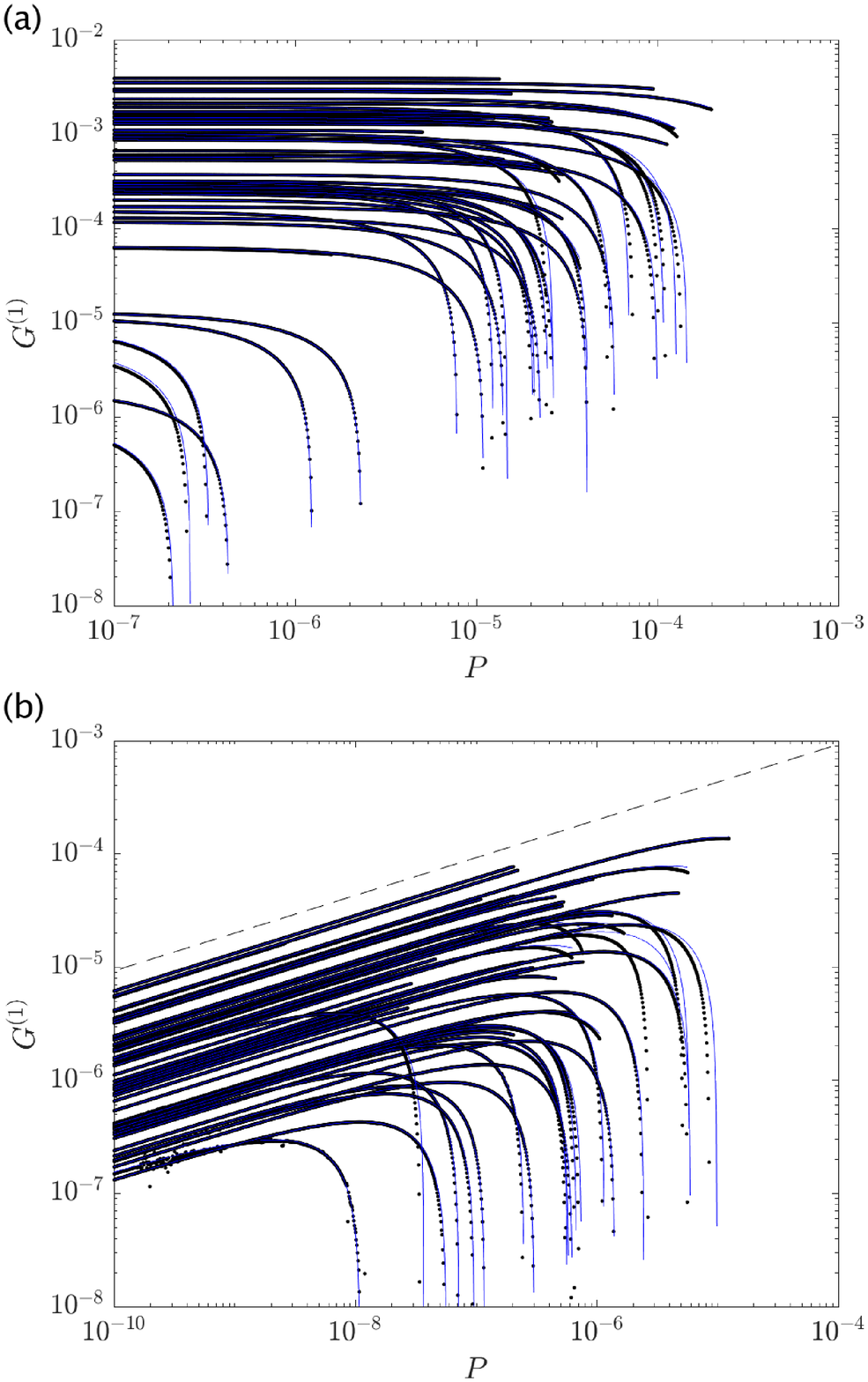}
    \caption{Shear modulus $G^{(1)}$ versus pressure $P$ for individual isostatic packings with $N=32$ disks within geometrical families that maintain their interparticle contact networks for purely repulsive (a) linear ($\alpha =2$) and (b) Hertzian ($\alpha=5/2$) spring interactions. Each panel includes $50$ geometrical families. The blue lines indicate best fits to Eq.~\ref{family} for each geometrical family. In this figure, we do not show packings with $G^{(1)} < 0$. In (b) the dashed line has slope equal to $1/3$.}
    \label{fig:isostatic_N_32_all}
\end{figure}

An important question concerning the mechanical response of jammed particle packings is understanding how the shear modulus $G$ increases with pressure $P$ when they are isotropically compressed above jamming onset~\cite{vanderwerf}.  Effective medium theory, which assumes only an affine response of the system, predicts that the shear modulus scales as $G \sim P^{1/3}$ for jammed packings of spheres with repulsive Hertzian spring interactions~\cite{emt} (i.e. $\alpha = 5/2$ in Eq.~\ref{potential}~\cite{book}), whereas experiments and simulations have shown that $G$ increases more strongly than $P^{1/3}$ for packings of Hertzian spheres~\cite{makse,makse2,roux}.  Other studies of sphere packings with repulsive linear spring interactions have shown that the ensemble-averaged shear modulus is constant at low pressures, $G(P) \sim P^{1/2}$ at high pressures, and the crossover pressure that separates the two scaling regimes decreases as $1/N^2$ with increasing system size~\cite{goodrich}.

What determines the pressure dependence of the shear modulus as packings are compressed above jamming onset?  We have shown in previous studies~\cite{vanderwerf} of sphere packings with repulsive linear spring interactions (i.e. $\alpha=2$ in Eq.~\ref{potential}) that the pressure dependence of the shear modulus is controlled by two key contributions: geometrical families~\cite{shen} and changes in the interparticle contact network~\cite{manning}.  For repulsive linear spring interactions, we find that the shear modulus $G^f$ varies continuously as $G^f = G^f_0 - B^f P$, where $G^f_0$ and $B^f$ are constants,
for a series of packings in a given geometrical family that differ in pressure, but possess the same network of interparticle contacts. A key finding of this prior work is that the shear modulus {\it decreases} with increasing pressure when the interparticle contact network does not change during compression. Geometrical families begin and end at ``point" and ``jump changes" in the contact network~\cite{manning,tuckman}. 
Point changes involve the addition or removal of a single interparticle contact (or multiple contacts when a rattler is added or removed from the contact network) without signficant particle motion. Point changes give rise to a discontontinuous jump in the shear modulus for power-law exponent $\alpha=2$ in Eq.~\ref{potential}, but not for $\alpha>2$.  In contrast, jump changes correspond to mechanical instabilities~\cite{lacks,maloney} with multiple simultaneous changes in the contact network and a discontinuous jump in the shear modulus across the jump change for any $\alpha$. At low pressures, where there are very few changes in the contact network, the geometrical family contribution dominates the ensemble-averaged shear modulus, and thus $G \sim G^f_0$ for sphere packings with repulsive linear spring interactions in the $P\rightarrow 0$ limit. At finite pressure, both geometrical families and changes in the contact network contribute to the pressure dependence of the ensemble-averaged shear modulus.  

In this article, using numerical simulations, we generalize the description of the pressure dependence of the shear modulus for packings of spherical particles compressed above jamming onset to systems with purely repulsive interactions and $\alpha > 2$, which includes packings of Hertzian spheres.  We find several important results. First, we show that the shear modulus for {\it isostatic} packings within geometrical families obeys
\begin{equation}
    \label{family}
    G^{(1)} = A^{(1)}_{\alpha} P^{(\alpha - 2)/(\alpha -1)} - B^{(1)}_{\alpha} P,
\end{equation}
where $A^{(1)}_{\alpha}$ and $B^{(1)}_{\alpha}$ depend only weakly on pressure. We decompose $G^{(1)}$ for each geometrical family into affine $G_a^{(1)}$ and non-affine $G^{(1)}_{n}$ contributions~\cite{mizuno,maloney,zaccone2011}, $G^{(1)} = G^{(1)}_a - G^{(1)}_{n}$, and show analytically that $G^{(1)}_a$ obeys the same form as Eq.~\ref{family}.  The isostatic geometrical family contribution to the shear modulus for Hertzian spheres has the form: $G^{(1)} = A^{(1)}_{5/2} P^{1/3} - B^{(1)}_{5/2} P$. The shear modulus $G^{(1)}$ first increases with increasing pressure and then in the absence of changes in the contact network, $G^{(1)}$ decreases due to the $-B^{(1)} P$ term, which can drive the system towards a mechanical instability with $G < 0$~\cite{tighe}. As shown previously for packings of spherical particles with repulsive linear spring interactions~\cite{vanderwerf}, we find that both geometrical families and changes in the contact network determine the scaling of the ensemble-averaged shear modulus at finite pressure for all $\alpha \ge 2$.  The ensemble-averaged shear modulus scales as $G(P) \sim P^a$, where $a \sim (\alpha - 2)/(\alpha -1)$  at low pressures below a characteristic pressure $P_c \sim 1/N^{2(\alpha-1)}$, and $G(P) \sim P^b$, where $b \sim (\alpha - 3/2)/(\alpha -1)$ for $P > P_c$. Specifically, for Hertzian spheres, we find that $G(P) \sim P^{1/3}$ for $P < P_c$ and $G(P) \sim P^{2/3}$ for $P > P_c$, which is consistent with prior experimental~\cite{emt} and simulation results~\cite{makse,jamming}. Based on these studies, we predict that the power-law scaling exponent for $\langle G(P) \rangle$ approaches unity and the crossover pressure tends to zero in the large-$\alpha$ limit.  

The remainder of this article is organized as follows. In Sec.~\ref{methods}, we describe the numerical methods used in this study, including the quasistatic, isotropic compression protocol used to generate the jammed packings and the calculations of the pressure, shear stress, and shear modulus for the jammed packings.  The key results are presented in Sec.~\ref{results}.  We first describe the calculations of the shear modulus as a function of pressure for packings in isostatic geometrical families with the same interparticle contact network.  We then determine analytically the affine contribution to the shear modulus within a given geometrical family and compare the affine shear modulus to the shear modulus obtained from sphere packings undergoing quasistatic, isostropic compression.  We calculate the ensemble-averaged shear modulus $\langle G \rangle$ as a function of pressure and show how the scaling of $\langle G \rangle$ with pressure varies with the power-law exponent $\alpha$ of the purely repulsive interparticle potential.  For each $\alpha$, we decompose $\langle G(P)\rangle$ into contributions from geometrical families and changes in the contact network and show that both contributions are important at finite pressure in the large-system limit. In Sec.~\ref{conclusions}, we summarize our conclusions and suggest future research directions, such as investigating the pressure-dependence of the shear modulus for packings in geometrical families at high pressures (not only for near-isostatic geometrical families), determining whether point and jump changes in the contact network occur for frictional packings using the ``bumpy-particle" model~\cite{bumpy}, and calculating the geometrical family and rearrangement contributions to the shear modulus for packings of nonspherical particles~\cite{shen}. We also include three appendices to provide additional technical details that supplement the descriptions in the main text. In Appendix A, we calculate the shear modulus for isostatic geometrical families as a function of pressure for disk packings with $\alpha=3$ and for sphere packings with $\alpha=2$ and $5/2$.  In Appendix B, we include a derivation of the decomposition of the shear modulus into the affine and non-affine terms and provide explicit expressions to calculate the non-affine term~\cite{maloney,zaccone2011}.  In Appendix C, we show that since the isostatic geometrical family contribution to the shear modulus includes a strongly negative term, the shear modulus can become negative for jammed packings generated at fixed shear strain~\cite{tighe}. 

\section{methods} 
\label{methods}

We investigate the mechanical properties of isotropically compressed jammed packings of bidisperse disks in 2D and spheres in 3D, containing $N/2$ large and $N/2$ small particles, each with the same mass $m$, and diameter ratio $\sigma_{l}/\sigma_{s}=1.4$. The particles are confined within a square/cubic box with side lengths, $L_{x}=L_{y}=1$ in 2D or $L_x=L_y=L_z=1$ in 3D, and periodic boundary conditions in all directions. We consider pairwise, purely repulsive, finite-ranged interactions between particles of the form in Eq.~\ref{potential}, for which the potential energy scales as a power-law in the overlap between pairs of particles with exponent $\alpha$. Pair forces are calculated using ${\vec f}_{ij} = dU(r_{ij})/dr_{ij} {\hat r}_{ij}$, where ${\hat r}_{ij} = {\vec r}_{ij}/r_{ij}$ is the unit vector vector pointing from the center of particle $j$ to the center of particle $i$. Results are presented below for $\alpha=2$ (linear springs), $5/2$ (Hertzian springs), and $3$. The pressure, shear stress, and shear modulus are expressed in units of $\epsilon/\sigma_s^d$ and forces are expressed in units of $\epsilon/\sigma_s$ below.   

We calculate the stress tensor ${\hat \Sigma}$ for each mechanically stable packing using the virial expression~\cite{bagi1996stress}:
\begin{equation}
    \label{stress_tensor}
    {\hat \Sigma}_{\beta \delta} = L^{-d} \sum_{i > j}f_{i j \beta}r_{i j \delta},
\end{equation}
where $\beta$, $\delta=x$, $y$, or $z$, $f_{i j \beta}$ is the $\beta$-component of the interparticle force ${\vec f}_{ij}$ on particle $i$ due to particle $j$, and $r_{ij\delta}$ is the $\delta$-component of the separation vector ${\vec r}_{ij}$. Note that we exclude rattler particles when calculating ${\hat \Sigma}_{\beta \delta}$. We define the shear stress as $\Sigma=-{\hat \Sigma}_{xy}$ and the pressure as $P = {\hat \Sigma}_{\beta \beta}/d$. To calculate the shear modulus $G$ numerically for each packing, we apply a series of small affine simple shear strain steps, $x_i' = x_i + d\gamma y_i$, to the packing in combination with Lees-Edwards boundary conditions, where $d\gamma = 10^{-9}$ is the shear strain increment, and minimize the total potential energy $U = \sum_{i >j} U(r_{ij})$ using the FIRE algorithm~\cite{bitzek2006structural} after each applied shear strain. We then measure $G = d\Sigma/d\gamma$ in the $\gamma \rightarrow 0$ limit. 

Below, we will characterize the shear modulus as a function of pressure from the onset of jamming near $P=0$ to systems that are significantly compressed with overlaps $\langle r_{ij} - \sigma_{ij}\rangle/\sigma_{ij} \approx 1\%$. To initialize the system, we randomly place particles in the simulation cell at rest and with no overlaps at packing fraction $\phi<0.01$. We increase the packing fraction in small increments $d\phi$ by increasing the particle diameters uniformly, and following each compression step, we minimize the total potential energy $U$. Energy minimization is terminated when $(\sum_i {\vec f}_i/N)^{2}<10^{-32}$, where ${\vec f}_i = \sum_j {\vec f}_{ij}$. Note that energy minimization can terminate when all of the pair forces $f_{ij}$ are near zero (i.e. the system is unjammed) or when the system achieves force balance. After each compression step, we measure the pressure $P$ and compare it to a target pressure $P_{t}$. If $P<P_{t}$, we compress the system by $d\phi$ and minimize the total potential energy. If $P>P_{t}$, we return to the system with the lower pressure, reduce the packing fraction increment from $d\phi$ to $d\phi/2$, and compress the system again, and repeat the process. This process is terminated when the pressure satisfies $P_{t}<P<(1+\zeta)P_{t}$, where $\zeta=10^{-7}$.

\begin{figure}[h]
    \centering
    \includegraphics[height=12cm]{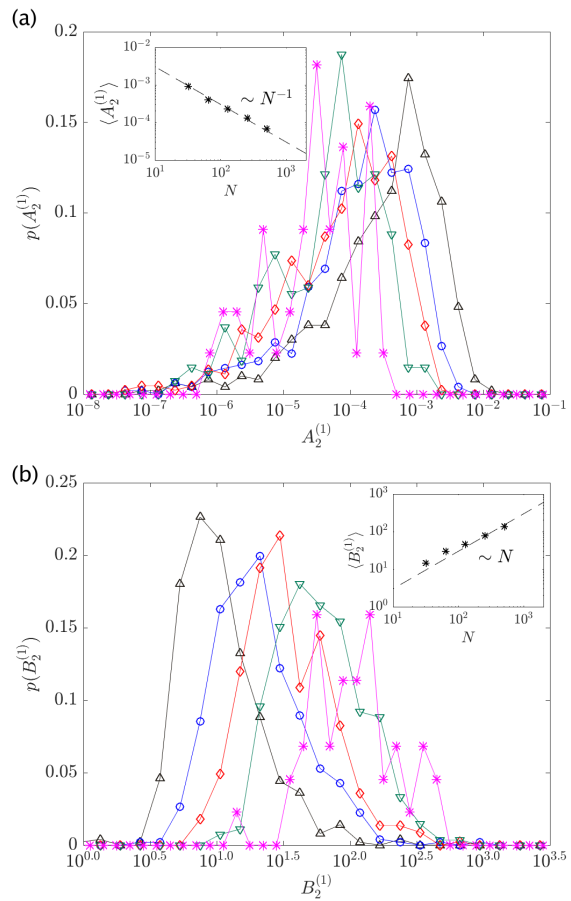}
    \caption{Probability distributions (a) $p(A^{(1)}_{2})$ and (b) $p(B^{(1)}_{2})$ of the coefficients $A^{(1)}_{2}$ and $B^{(1)}_{2}$ in Eq.~\ref{family} for disk packings interacting via repulsive linear spring forces within isostatic geometrical families for $N=32$ (black upward triangles), $64$ (blue circles), $128$ (red diamonds), $256$ (cyan downward triangles), and $512$ (magenta asterisks). The insets to panels (a) and (b) display $\langle A^{(1)}_2 \rangle$ and $\langle B^{(1)}_2 \rangle$ (averaged over geometrical families) versus system size $N$, respectively. The dashed lines have slopes equal to $-1$ and $1$ in the insets to panels (a) and (b).}
    \label{fig:isostatic_N_all_A_B_distribution_linear}
\end{figure}

\begin{figure}[h]
    \centering
    \includegraphics[height=12cm]{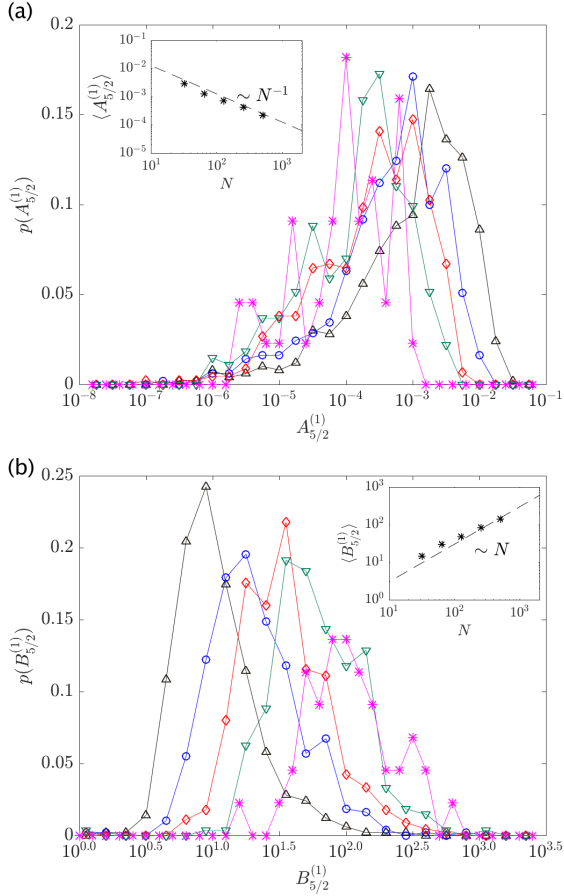}
    \caption{Probability distributions (a) $p(A^{(1)}_{5/2})$ and (b) $p(B^{(1)}_{5/2})$ of the coefficients $A^{(1)}_{5/2}$ and $B^{(1)}_{5/2}$ in Eq.~\ref{family} for disk packings interacting via repulsive Hertzian spring forces within isostatic geometrical families for $N=32$ (black upward triangles), $64$ (blue circles), $128$ (red diamonds), $256$ (cyan downward triangles), and $512$ (magenta asterisks). The insets to panels (a) and (b) display $\langle A^{(1)}_{5/2} \rangle$ and $\langle B^{(1)}_{5/2} \rangle$ (averaged over geometrical families) versus system size $N$, respectively. The dashed lines have slopes equal to $-1$ and $1$ in the insets to panels (a) and (b).}
    \label{fig:isostatic_N_all_A_B_distribution_Hertzian}
\end{figure}

We sample more than $1000$ jammed packings logarithmically in pressure, spanning from isostatic packings at $P = 10^{-7}$ to compressed states with $P=10^{-2}$ for $\alpha =2$. To generate packings of particles interacting via Eq.~\ref{potential} with $\alpha=5/2$ and $3$, we initialized the system with isostatic packings generated using $\alpha=2$ and then performed the compression protocol using the appropriate $\alpha$. With this initialization, we ensure that the isostatic contact networks are the same for all $\alpha$ that we studied.  For $\alpha=5/2$ and $3$, the pressures that we sample vary from $P=10^{-10}$ to $10^{-2}$. 

\begin{figure}[h]
    \centering
    \includegraphics[height=12cm]{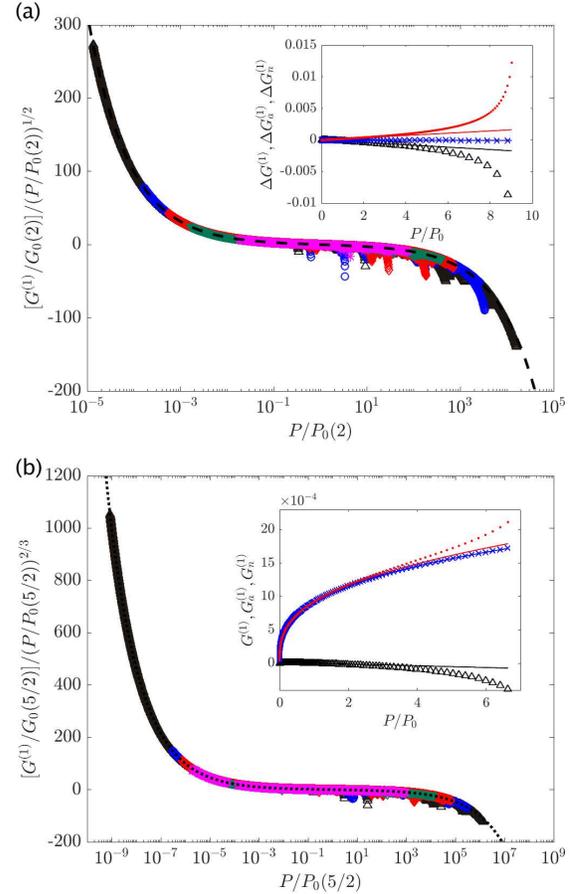}
    \caption{(a) $(G^{(1)}/G_0) (P/P_0)^{-1/2}$ and (b) $(G^{(1)}/G_0) (P/P_0)^{-2/3}$ for isostatic geometrical families plotted versus $P/P_0$ for disk packings with repulsive linear ($\alpha=2$) and Hertzian ($\alpha=5/2$) spring interactions, respectively, for several system sizes: $N=32$ (black upper triangles), $64$ (blue circles), $128$ (red diamonds), $256$ (green downward triangles), and $512$ (magenta asterisks). The dashed lines in (a) and (b) are plots of Eq.~\ref{dimensionless2} for $\alpha=2$ and $5/2$, respectively. In the insets to panels (a) and (b), we show $\Delta G^{(1)} = G^{(1)}(P/P_0) - G^{(1)}(0)$ or $G^{(1)}(P/P_0)$ (black upward triangles), $\Delta G^{(1)}_{n} = G^{(1)}_n(P/P_0) - G^{(1)}_n(0)$ or $G^{(1)}_n(P/P_0)$ (red dots), and $\Delta G^{(1)}_{a} = G^{(1)}_a(P/P_0) - G^{(1)}_a(0)$ or $G^{(1)}_a(P/P_0)$ (blue exes) with best fits to Eq.~\ref{family} (black, red, and blue solid lines, respectively) for an example $N=32$ packing with $\alpha =2$ (inset to (a)) and $5/2$ (inset to (b)).}
    \label{fig:isostatic_collapse}
\end{figure}

\begin{figure}[h]
    \centering
    \includegraphics[height=12cm]{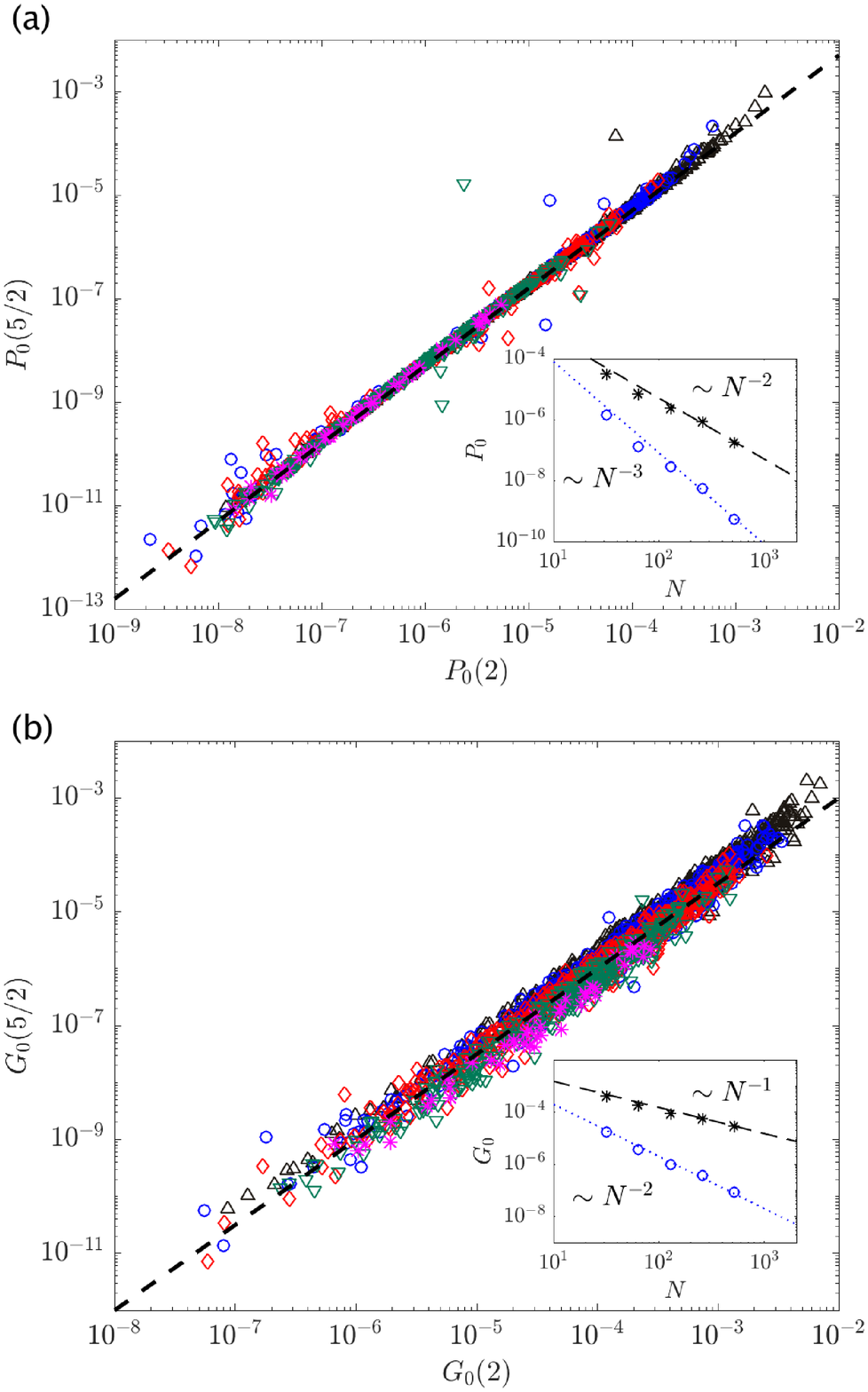}
    \caption{(a) $P_{0}(5/2)$ versus $P_0(2)$ and (b) $G_{0}(5/2)$ versus $G_0(2)$ for disk packings with repulsive linear ($\alpha=2$) and Hertzian ($\alpha=5/2$) spring interactions for several system sizes: $N=32$ (black upper triangles), $64$ (blue circles), $128$ (red diamonds), and $256$ (cyan downward triangles), and $512$ (magenta asterisks).  The dashed lines have slopes equal to $3/2$. The insets display the system size dependence of $P_0$ and $G_0$ for $\alpha=2$ (black asterisks) and $5/2$ (blue circles). The dashed and dotted lines in the inset to panel (a) have slopes equal to $-2$ and $-3$, respectively. The dashed and dotted lines in the inset to panel (b) have slopes equal to $-1$ and $-2$, respectively.}
    \label{fig:isostatic_G0_P0}
\end{figure}

\section{Results} 
\label{results}

The results concerning the mechanical properties of jammed packings of spherical particles with finite-ranged, purely repulsive interactions will be presented in four sections below. In Sec.~\ref{results:isostatic_families}, we investigate the shear modulus $G^{(1)}$ for isostatic packings of spherical particles that occur in individual geometrical families (for power-law exponents $\alpha = 2$, $5/2$, and $3$ and several system sizes) and determine how $G^{(1)}$ varies with pressure prior to the first change in the interparticle contact network.  In Sec.~\ref{results:affine_approximation}, we decompose $G^{(1)}$ for isostatic geometrical families into the affine and non-affine contributions. We show that the non-affine contribution plays an important role in determining the behavior of $G^{(1)}(P)$ even though the contact network does not change.  In Sec.~\ref{results:second}, we measure the shear modulus of packings during isotropic compression as they undergo point or jump changes in the contact network and the system transitions from the first, isostatic geometrical family to the second geometrical family. In Sec.~\ref{results:rearrangement_contribution}, we calculate the ensemble-averaged shear modulus $\langle G \rangle$ and find a master curve for $\langle G(P)\rangle$ as a function of system size. The master curve is not simply a sum of two power-laws, but $\langle G(P) \rangle \sim P^a$, where $a \sim (\alpha - 2)/(\alpha -1)$, below a characteristic pressure $P_c \sim 1/N^{2(\alpha-1)}$, and $\langle G(P) \rangle \sim P^b$, where $b \sim (\alpha - 3/2)/(\alpha -1)$, for $P > P_c$.  To better understand the scaling behavior, we decompose $\langle G\rangle =\langle G_f \rangle+ \langle G_r\rangle$ into contributions from geometrical families $G_{f}$ and changes in the contact network $G_r$.  We show that in the large-system limit both contributions are important for determining the ensemble-averaged shear modulus $\langle G \rangle$ at finite pressure. 

\subsection{Isostatic Geometrical Families}
\label{results:isostatic_families}

\begin{figure}[h]
    \centering
    \includegraphics[height=12cm]{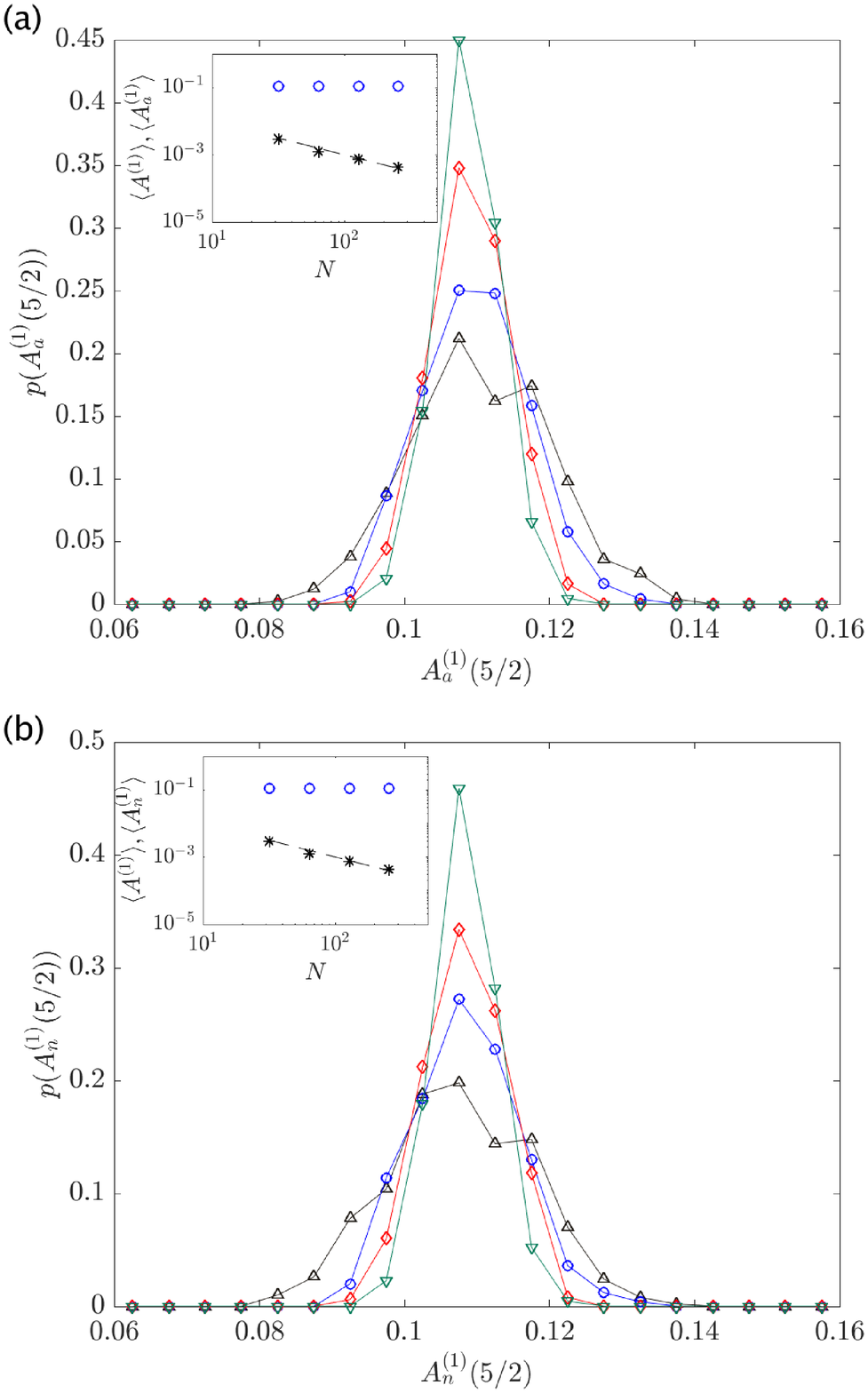}
    \caption{The probability distributions of the coefficients (a) $A^{(1)}_{a}(5/2)$ and (b) $A^{(1)}_{n}(5/2)$ that contribute to the shear modulus $G^{(1)}$ of isostatic geometrical families for $N=32$ (black upward triangles), $64$ (blue circles), $128$ (red diamonds), and $256$ (green downward triangles) disk packings with repulsive Hertzian spring interactions.  $G^{(1)} = A^{(1)}_{5/2} P^{1/3} -B^{(1)}_{5/2} P$, and $A^{(1)}_{5/2}=A^{(1)}_a(5/2) - A^{(1)}_n(5/2)$, where $A^{(1)}_a(5/2)$ and $A^{(1)}_n(5/2)$ are the affine and non-affine 
    contributions to the $P^{1/3}$ term for $G^{(1)}$, respectively. The insets in (a) and (b) show the average values of $A^{(1)}_a$, $A^{(1)}_n$, and $A^{(1)}$ versus system size $N$.}
    \label{fig:Aaffine_Anonaffine_N_all}
\end{figure}

\begin{figure}[h]
    \centering
    \includegraphics[height=12cm]{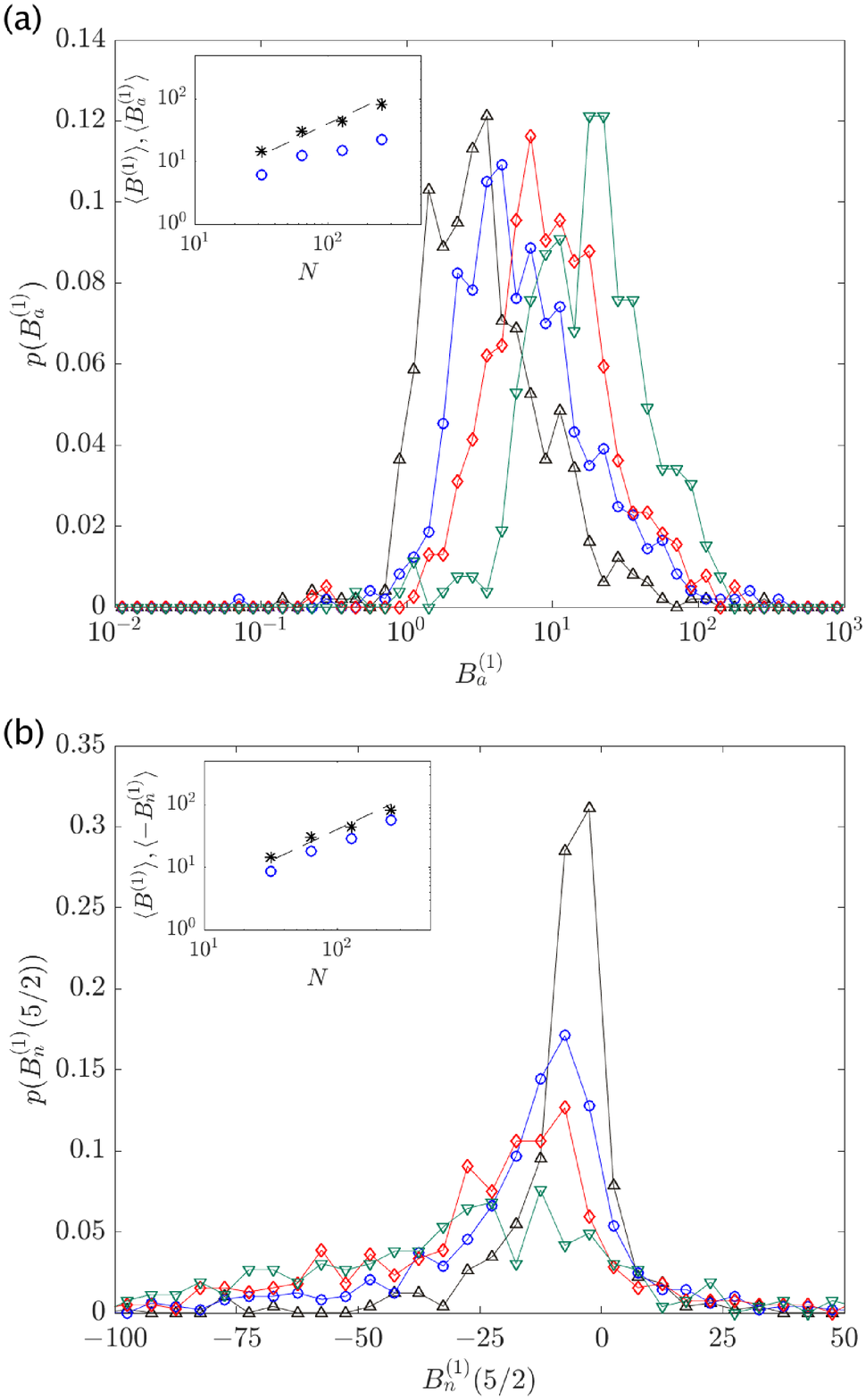}
    \caption{The probability distributions of the coefficients (a) $B^{(1)}_{a}$ and (b) $B^{(1)}_{n}(5/2)$ that contribute to the shear modulus $G^{(1)}$ of isostatic geometrical families for $N=32$ (black upward triangles), $64$ (blue circles), $128$ (red diamonds), and $256$ (green downward triangles) disk packings with repulsive Hertzian spring interactions.  $G^{(1)} = A^{(1)}_{5/2} P^{1/3} -B^{(1)}_{5/2} P$, and $B^{(1)}_{5/2}=B^{(1)}_a - B^{(1)}_n(5/2)$, where $B^{(1)}_a$ and $B^{(1)}_n(5/2)$ are the affine and non-affine 
    contributions to the $-P$ term for $G^{(1)}$, respectively. The insets in (a) and (b) show the average values of $B^{(1)}_a$, $B^{(1)}_n$, and $B^{(1)}$ versus system size $N$.}
    \label{fig:Baffine_Bnonaffine_N_all}
\end{figure}

Isotropically compressed jammed packings occur as geometrical families as a function of pressure.  Specifically, if we consider a packing at jamming onset with $P=0$, it will possess packing fraction $\phi_J$, non-rattler particle positions ${\vec R} = \{x_1,x_2,\ldots,x_N,y_1,y_2,\ldots,y_N\}$ in 2D or 
${\vec R} = \{x_1,x_2,\ldots,x_N,y_1,y_2,\ldots,y_N,z_1,z_2,\ldots,z_N\}$ in 3D, and a contact network with an isostatic number of contacts, $N_c=N_c^0$.  If we compress the jammed system by $d\phi$ (and minimize the total potential energy), the particle positions will change continuously with $d\phi$ to ${\vec R}'$, the pressure will become nonzero, and as long as $d\phi$ is sufficiently small, the interparticle contact network will not change. At a given pressure $P^*$, which is different for each isostatic contact network, the contact network will undergo a point change or a jump change~\cite{tuckman}.  In Fig.~\ref{fig:contactnetworkChange}, we show the contact network for an isostatic jammed packing (with $N=32$ and $\alpha=2$) near the onset of jamming (with $P=10^{-7}$) and immediately before (with $P=4.60 \times 10^{-6}$) and after (with $P=4.65 \times 10^{-6}$) a point change. After the point change, the jammed packing has one extra interparticle contact and $N_c = N_c^0+1$.  This behavior is similar for isotropically compressed packings with larger system sizes, except $P^*$ decreases with increasing system size.  

In Fig.~\ref{fig:isostatic_N_32_all} (a), we show the shear modulus $G^{(1)}$ for all isostatic disk packings within each of $50$ different geometrical families generated using $N=32$ and repulsive linear spring interactions ($\alpha=2$).  We find that for each geometrical family, $G^{(1)}$ tends to a constant in the $P\rightarrow 0$ limit, and decreases strongly with increasing pressure.  (We find similar results for sphere packings in 3D with repulsive linear spring interactions as shown in Fig.~\ref{fig:G_P_3D_N_64} (a) in Appendix A.) We fit the shear modulus for each geometrical family to $G^{(1)}(P) = A^{(1)}_2 - B^{(1)}_2 P$ (Eq.~\ref{family}) and show the results for the coefficients $A^{(1)}_2$ and $B^{(1)}_2$ for several system sizes in Fig.~\ref{fig:isostatic_N_all_A_B_distribution_linear}. We find that $\langle A^{(1)}_2\rangle \sim N^{-1}$ and $\langle B^{(1)}_2\rangle \sim N$ when averaged over all of the isostatic geometrical families, as shown in the insets to Fig.~\ref{fig:isostatic_N_all_A_B_distribution_linear} (a) and (b).

\begin{figure}[h]
    \centering
    \includegraphics[height=12cm]{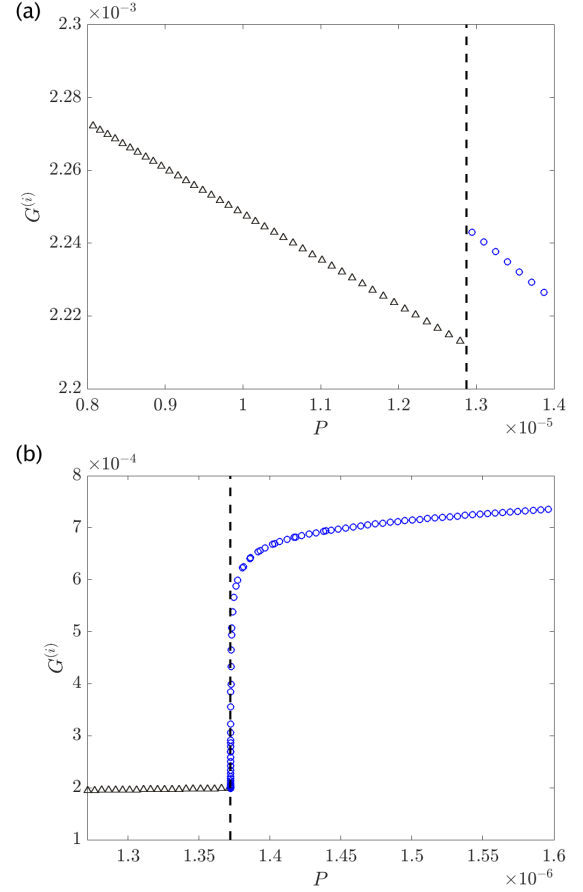}
    \caption{Shear modulus $G^{(i)}$ of a series of $N=32$ disk packings with repulsive (a) linear and (b) Hertzian spring interactions as the system undergoes a point change during isotropic compression (at $P \approx 1.29 \times 10^{-5}$ for $\alpha =2$ and $P=1.37 \times 10^{-6}$ for $\alpha=5/2$ indicated by vertical dashed lines) from the isostatic (black upward triangles) geometrical family with $N_c^0$ contacts to the second geometrical family (blue circles) with $N_c^0+1$ contacts.}
    \label{fig:G_P_second_family_N_32}
\end{figure}

To investigate the geometrical families that occur for jammed packings of repulsive Hertzian disks, we start with an isostatic disk packing generated using repulsive linear spring interactions at the lowest pressure we considered, change the interaction potential from $\alpha=2$ to $5/2$, and minimize the total potential energy.  We verified that each lowest-pressure, isostatic packing for repulsive linear spring interactions gives rise to an isostatic packing for repulsive Hertzian spring interactions. We then repeat (for repulsive Hertzian spring interactions) the same isotropic compression protocol used to generate isostatic geometrical families for particles with repulsive linear spring interactions. 
We show the shear modulus $G^{(1)}$ for the isostatic geometrical families for repulsive Hertzian disks in Fig.~\ref{fig:isostatic_N_32_all} (b).  In contrast to the results for repulsive linear spring interactions, $G^{(1)} \rightarrow 0$ in the $P \rightarrow 0$ limit.  As we found for $\alpha=2$, $G^{(1)}$ also decreases at sufficiently large pressures. (We find similar results for sphere packings in 3D with repulsive Hertzian spring interactions as shown in Fig.~\ref{fig:G_P_3D_N_64} (b) in Appendix A.) We fit each geometrical family to $G^{(1)} = A^{(1)}_{5/2} P^{1/3} -B^{(1)}_{5/2} P$ (Eq.~\ref{family} with $\alpha=5/2$) for several system sizes and show the results for the coefficients $A^{(1)}_{5/2}$ and $B^{(1)}_{5/2}$ in Fig.~\ref{fig:isostatic_N_all_A_B_distribution_Hertzian}. As we found for $\alpha=2$, $\langle A^{(1)}_{5/2}\rangle \sim N^{-1}$ and $\langle B^{(1)}_{5/2}\rangle \sim N$ when averaged over all of the isostatic geometrical families, which is verified in the insets to Fig.~\ref{fig:isostatic_N_all_A_B_distribution_Hertzian} (a) and (b). 

\begin{figure}[h]
    \centering
    \includegraphics[height=12cm]{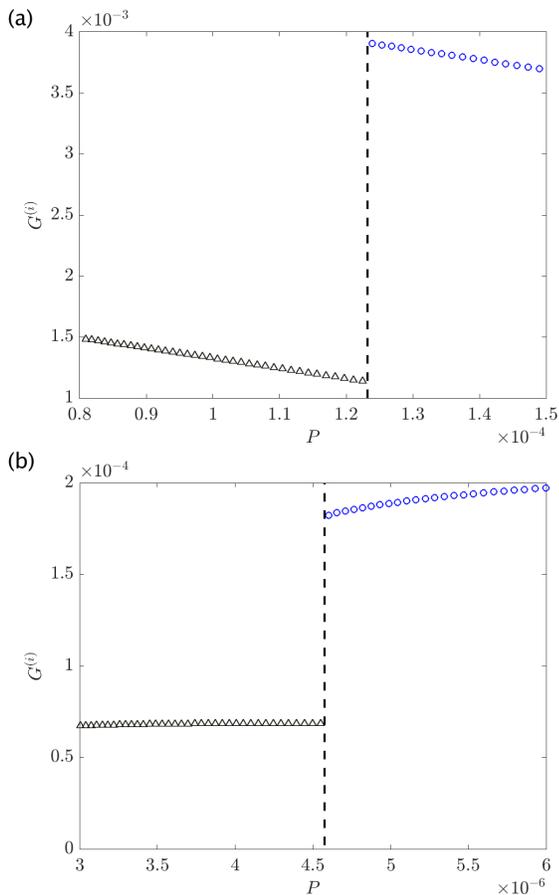}
    \caption{Shear modulus $G^{(i)}$ of a series of $N=32$ disk packings with repulsive (a) linear and (b) Hertzian spring interactions as the system undergoes a jump change during isotropic compression (at $P \approx 1.23 \times 10^{-4}$ for $\alpha =2$ and $P=4.57 \times 10^{-6}$ for $\alpha=5/2$ indicated by vertical dashed lines) from the isostatic (black upward triangles) geometrical family with $N_c^0$ contacts to a second geometrical family (blue circles) with $N_c^0+1$ contacts.}
    \label{fig:G_P_second_family_N_32_jump}
\end{figure}

The form for $G^{(1)}(P)$ in Eq.~\ref{family} is motivated by the affine contribution to $G^{(1)}$, which can be calculated analytically as discussed below in Sec.~\ref{results:affine_approximation}.  Using Eq.~\ref{family}, we predict $G^{(1)} = A^{(1)}_3 P^{1/2} - B^{(1)}_3 P$ for disk packings with $\alpha=3$, which is verified in Fig.~\ref{fig:G_P_N_32_alpha_3} in Appendix A. Thus, from Eq.~\ref{family}, we find that $G^{(1)}(P)$ is well-defined for disk packings with repulsive interactions with $\alpha \ge 2$; $G^{(1)}(P)$ tends to zero in the $P\rightarrow 0$ limit for all $\alpha >2$; and after a characteristic pressure that depends on the power-law exponent $\alpha$ and system size $N$, $G^{(1)}(P)$ decreases linearly with increasing pressure for all $\alpha$.  The $-B^{(1)}_{\alpha} P$ term can give rise to unstable packings with $G^{(1)} < 0$ at finite pressures~\cite{lakes,tighe,goodrich2}, but our results emphasize that all jammed packings possess $G^{(1)} >0$ at sufficiently low pressures. (See Appendix B for statistics of $G^{(1)} <0$ as a function of pressure and system size for several $\alpha$ values.)  

\begin{figure*}
    \centering
    \includegraphics[height=4.75cm]{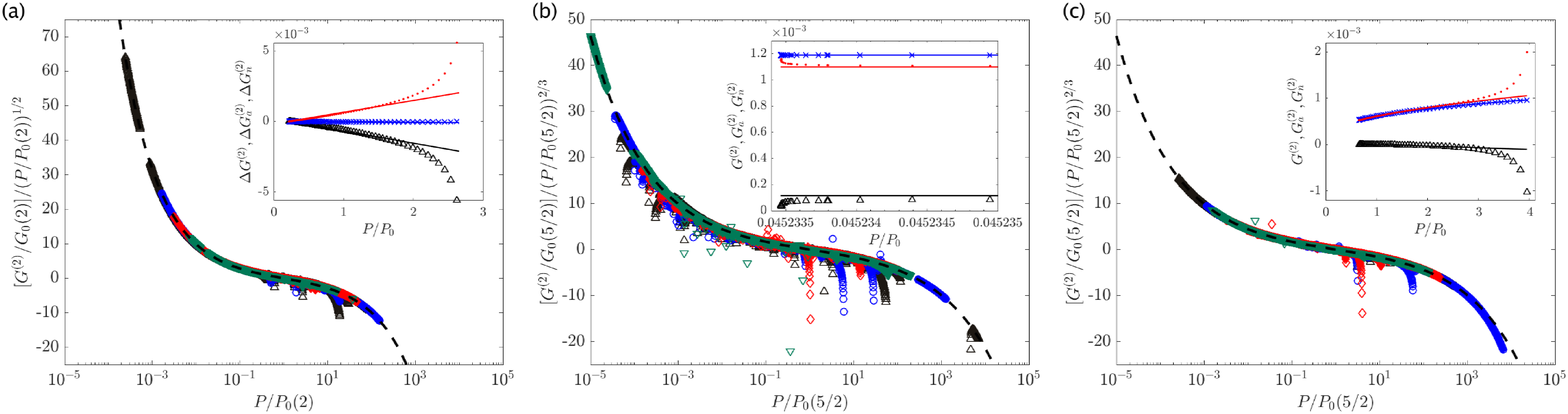}
    \caption{(a) $(G^{(2)}/G_0) (P/P_0)^{-1/2}$ versus $P/P_0$ for packings with $\alpha=2$ in the second geometrical family following a point or jump change in the contact network. $(G^{(2)}/G_0) (P/P_0)^{-2/3}$ is plotted versus $P/P_0$ for packings with $\alpha=5/2$ in the second geometrical family following (b) a point change or (c) a jump change in the contact network. In (a)-(c), several system sizes are shown: $N=32$ (black upper triangles), $64$ (blue circles), $128$ (red diamonds), $256$ (green downward triangles), and $512$ (magenta asterisks). The dashed lines in (a)-(c) are plots of Eq.~\ref{dimensionless2} for $\alpha=2$ in (a) and $5/2$ in (b) and (c).  In the insets to panels (a)-(c), we show $\Delta G^{(1)} = G^{(1)}(P/P_0) - G^{(1)}(0)$ or $G^{(1)}(P/P_0)$ (black upward triangles), $\Delta G^{(1)}_{n} = G^{(1)}_n(P/P_0) - G^{(1)}_n(0)$ or $G^{(1)}_n(P/P_0)$ (red dots), and $\Delta G^{(1)}_{a} = G^{(1)}_a(P/P_0) - G^{(1)}_a(0)$ or $G^{(1)}_a(P/P_0)$ (blue exes) with best fits to Eq.~\ref{family} (black, red, and blue solid lines, respectively) for an example $N=32$ packing with $\alpha =2$ (inset to (a)) and $5/2$ (inset to (b)).}
    \label{fig:2nd_families_collapse}
\end{figure*}

We can rewrite Eq.~\ref{family} in dimensionless form, 
\begin{equation}
    \label{dimensionless}
    \frac{G^{(1)}}{G_0(\alpha)} = \left( \frac{P}{P_0(\alpha)}  \right)^{\frac{\alpha-2}{\alpha-1}} -\frac{P}{P_0(\alpha)}    
\end{equation}
to assess the quality of the data collapse for the shear modulus of isostatic geometrical families, where $G_0(\alpha)=(A^{(1)}_{\alpha})^{\alpha-1} (B^{(1)}_{\alpha})^{2-\alpha}$ and $P_0(\alpha)=(A^{(1)}_{\alpha}/B^i_{\alpha})^{\alpha-1}$.  We can then multiply both sides of Eq.~\ref{dimensionless} by $(P/P_0(\alpha))^{-(\alpha-3/2)/(\alpha-1)}$ to yield the symmetric form: 
\begin{eqnarray}
    \label{dimensionless2}
    \frac{G^{(1)}}{G_0(\alpha)} \left( \frac{P}{P_0(\alpha)} \right)^{-\frac{\alpha-3/2}{\alpha-1}} & = & \left( \frac{P}{P_0(\alpha)} \right)^{-\frac{1}{2(\alpha-1)}} \nonumber \\
    & - & \left( \frac{P}{P_0(\alpha)} \right)^{\frac{1}{2(\alpha-1)}}.
\end{eqnarray}
$P_0(\alpha)$ is the crossover pressure that separates the two power-law scaling behaviors in Eq.~\ref{dimensionless2}. For $P \gg P_0(\alpha)$, the term on the right hand side of Eq.~\ref{dimensionless2} with the positive exponent will dominate, whereas for $P \ll P_0(\alpha)$, the term with the negative exponent will dominate.  In Fig.~\ref{fig:isostatic_collapse} (a) and (b), we plot $(G^i/G_0(\alpha)) (P/P_0(\alpha))^{-(\alpha-3/2)/(\alpha-1)} $ versus $P/P_0(\alpha)$. The data for $G^{(1)}$ shows reasonable collapse onto a master curve for the shear modulus for repulsive linear ($\alpha=2)$ and Hertzian ($\alpha=5/2$) spring interactions for all isostatic packings that we generated.  For packings with repulsive linear spring interactions, the data for $(G^{(1)}/G_0(2)) (P/P_0(2))^{-1/2}$ for $P \ll P_0$ obeys power-law scaling with exponent $-1/2$ and for $P \gg P_0$ the data obeys power-law scaling with exponent $1/2$. For packings with repulsive Hertzian spring interactions, the data for $(G^{(1)}/G_0(5/2)) (P/P_0(5/2))^{-2/3}$ for $P \ll P_0$ obeys power-law scaling with exponent $-1/3$ and for $P \gg P_0$ the data obeys power-law scaling with exponent $1/3$.
Using Eq.~\ref{dimensionless}, we can show that the characteristic pressures $P_0$ and shear moduli $G_0$ for packings with repulsive linear and Hertzian spring interactions obey scaling relations: $P_0(5/2) \sim (P_0(2))^{3/2}$ and $G_0(5/2) \sim (G_0(2))^{3/2}$. (See Fig.~\ref{fig:isostatic_G0_P0}.) In addition, using Eq.~\ref{dimensionless}, we can show that $P_0 \sim N^{-2(\alpha-1)}$ and $G_0 \sim N^{-2(\alpha-3/2)}$ tend to zero in the large-system limit, which is verified in the simulation data shown in the insets to Fig.~\ref{fig:isostatic_G0_P0}.

In Fig.~\ref{fig:isostatic_collapse} (a) and (b), we notice that $G^{(1)}$ deviates from the dimensionless scaling form in Eq.~\ref{dimensionless2} at large $P/P_0$ for some of the packings with $\alpha =2$ and $5/2$.  In the insets to Fig.~\ref{fig:isostatic_collapse} (a) and (b), we show that $G^{(1)}$ for an example $N=32$ packing obeys the scaling form for pressures where {\it both} the affine and non-affine contributions to the shear modulus follow Eq.~\ref{dimensionless2}. The deviations of $G^{(1)}$ from the scaling form are caused by the growing non-affine contribution to the shear modulus. These results are the same for $G^{(1)}$ for all packings that possess deviations at large $P/P_0$ in Fig.~\ref{fig:isostatic_collapse} (a) and (b). (See Sec.~\ref{results:affine_approximation} below.) 
Since the non-affine motion is increasing toward the end of the isostatic geometrical family, it is likely that it is correlated with a mechanical instability~\cite{maloney,zaccone2011} arising from the change in the contact network from the first to the second geometrical family.  

\begin{figure}[h]
    \centering
    \includegraphics[height=6cm]{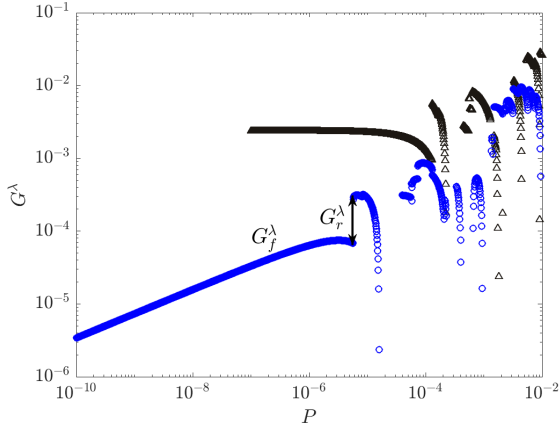}
    \caption{Shear modulus $G^{\lambda}$ for initial condition $\lambda$ at $P=0$ undergoing isotropic compression as a function of pressure $P$ for an $N=32$ packing with repulsive (a) linear (black upward triangles) and (b) Hertzian spring interactions (blue circles). $G^{\lambda}=G^{\lambda}_f+G^{\lambda}_r$ can be decomposed into the contributions from the continuous geometrical families $G^{\lambda}$ and discontinuities $G^{\lambda}_r$ caused by point and jump changes in the contact network.}
    \label{fig:G_P_entire_single_all_N_32}
\end{figure}

\begin{figure}[h]
    \centering
    \includegraphics[height=12cm]{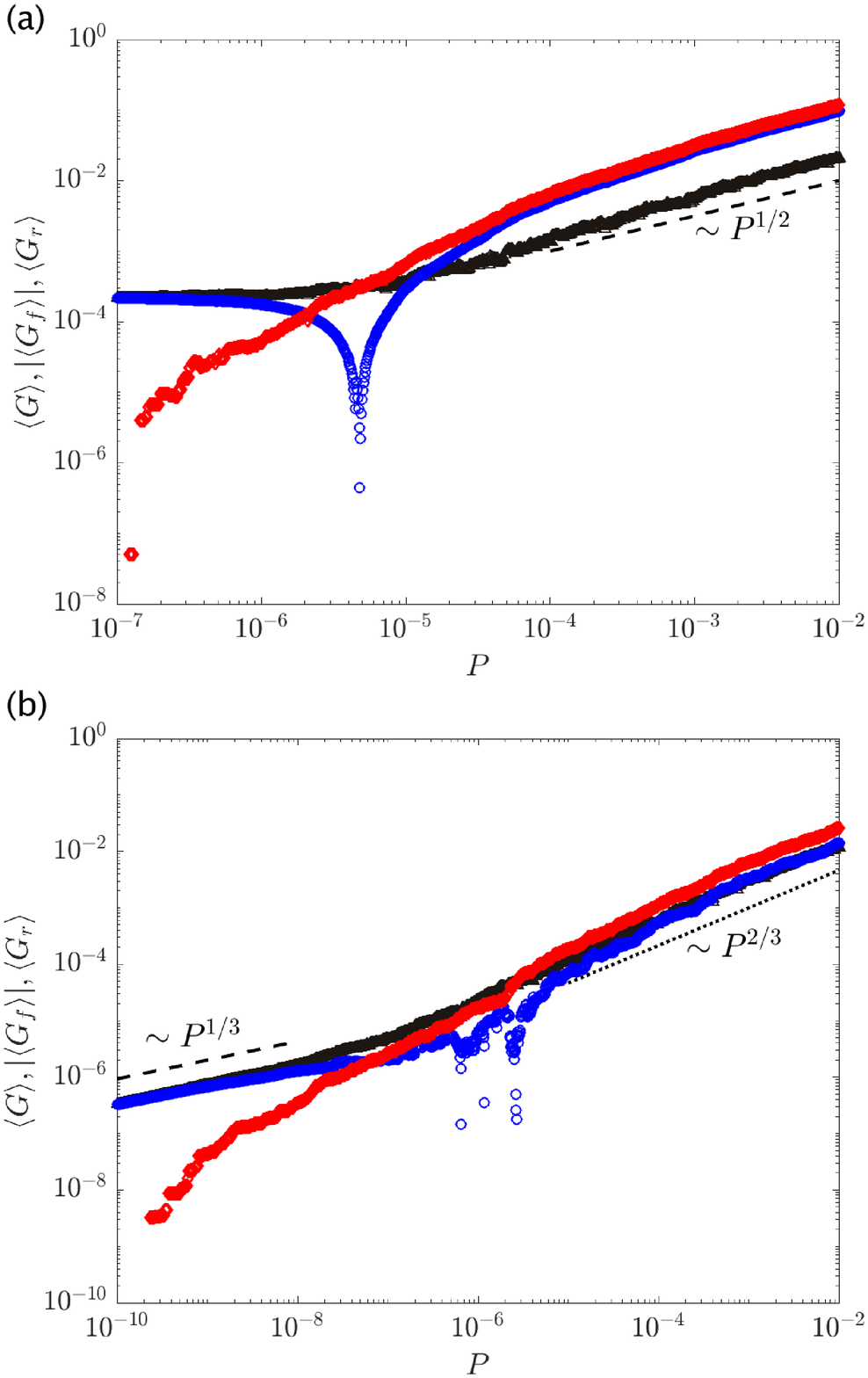}
    \caption{Ensemble-averaged shear modulus $\langle G \rangle$ (black upward triangles) as a function of pressure $P$ for $N=128$ packings with (a) $\alpha=2$ and (b) $5/2$ decomposed into contributions from geometrical families $| \langle G_f \rangle |$ (blue circles) and changes in the contact network $\langle G_{r} \rangle$ (red diamonds). In (a), the dashed line has slope equal to $1/2$ and in (b), the dashed and dotted lines have slopes equal to $1/3$ and $2/3$, respectively.}
    \label{fig:decouplingDemo}
\end{figure}

\subsection{Affine and Non-affine Contributions to $G^{(1)}$ }
\label{results:affine_approximation}

In the previous section, we calculated the pressure-dependent shear modulus for isostatic geometrical families at sufficiently low pressures such that the packings do not undergo point or jump changes in the contact network and showed that $G^{(1)}(P)$ obeys Eqs.~\ref{family} and~\ref{dimensionless} over a range of pressure.  In this section, we decompose $G^{(1)}=G^{(1)}_a - G^{(1)}_n$ into the affine $G^{(1)}_a$ and non-affine $G^{(1)}_n$ contributions and determine their relative magnitudes as a function of pressure~\cite{maloney,zaccone2011}.  The affine contribution considers the response of the packing to an ideal simple shear deformation without relaxation and assumes no changes in the interparticle contact network. In contrast, the non-affine contribution includes particle motion from energy minimization and, when we include transitions between geometrical families, the effects of changes in the interparticle contact network. (See Appendix B for a derivation of the non-affine contribution to the shear modulus~\cite{maloney,zaccone2011}.) 

We first focus on the affine contribution $G^{(1)}_a$ to the shear modulus of isostatic geometrical families. We consider packings near jamming onset and apply an affine simple shear shear deformation to their particle coordinates, $(x_i',y_i',z_i')=(x_i^0 + \gamma y_i^0,y_i^0,z_i^0)$ in 3D or $(x_i',y_i')=(x_i^0 + \gamma y_i^0,y_i^0)$ in 2D, where $(x_i^0,y_i^0,z_i^0)$ in 3D and $(x_i^0,y_i^0)$ in 2D are the particle positions in the original jammed packing, consistent with Lees-Edwards boundary conditions for simple shear strain $\gamma$. The affine contribution is obtained by calculating $G^{(1)}_a = \partial \Sigma/\partial \gamma$, where the shear stress is given by
\begin{eqnarray}
\label{stress2}
\Sigma &  = & L^{-d} \frac{\partial U}{\partial \gamma} \nonumber \\
& = &\epsilon L^{-d} \sum_{i>j} \left(-\frac{x_{ij}y_{ij}}{\sigma_{ij}r_{ij}}\right)\left(1-\frac{r_{ij}}{\sigma_{ij}}\right)^{\alpha-1} \times \nonumber \\ 
& & \Theta \left( 1- \frac{r_{ij}}{\sigma_{ij}} \right),
\end{eqnarray}
after inserting Eq.~\ref{potential} into Eq.~\ref{stress_tensor}. In Eq.~\ref{stress2}, $x_{ij}$ and $y_{ij}$ are the $x$- and $y$-separations between the centers of particles $i$ and $j$. Thus, for the affine contribution, $G^{(1)}_a =L^{-d} \partial^{2} U/\partial \gamma^{2}$, we obtain
\begin{equation}
\label{G_affine}
\begin{aligned}
G^{(1)}_{a} & = & L^{-d} \epsilon\sum_{i>j} \left[ -\left(1-\frac{r_{ij}}{\sigma_{ij}}\right)^{\alpha-1}\frac{y_{ij}^4}{\sigma_{ij}r_{ij}^3}+ \right . \\
& & \left. \left(\alpha-1\right)\frac{x_{ij}^2y_{ij}^2}{\sigma_{ij}^2r_{ij}^2}\left(1-\frac{r_{ij}}{\sigma_{ij}}\right)^{\alpha-2} \right] \Theta \left( 1- \frac{r_{ij}}{\sigma_{ij}} \right).
\end{aligned}
\end{equation}

To determine $G^{(1)}_a$ as a function of pressure, we write the pressure 
\begin{equation}
\label{pressure}
P= {\hat \Sigma}_{\beta \beta}/d = \frac{\epsilon}{dL^d} \sum_{i>j} \frac{r_{ij}}{\sigma_{ij}}\left(1-\frac{r_{ij}}{\sigma_{ij}}\right)^{\alpha-1} \Theta \left( 1- \frac{r_{ij}}{\sigma_{ij}} \right)
\end{equation}
in terms of the particle separations. If we define 
\begin{equation}
\label{definition}
{\overline P}_{ij}=\epsilon\frac{r_{ij}}{\sigma_{ij}}\left(1-\frac{r_{ij}}{\sigma_{ij}}\right)^{\alpha-1} \Theta \left( 1- \frac{r_{ij}}{\sigma_{ij}} \right)
\end{equation}
and assume that ${\overline P}_{ij}$ scales linearly with pressure for all $i$, $j$ pairs,
\begin{equation}
\label{assumption}
{\overline P}_{ij} = d L^d \chi_{ij} P, 
\end{equation}
where $\chi_{ij}$ is independent of pressure, we can use Eqs.~\ref{definition} and~\ref{assumption} to express $G^{(1)}_a$ in Eq.~\ref{G_affine} as a function of pressure. (We have verified numerically for packings with $\alpha=2$ and $5/2$ that $\chi_{ij}$ is nearly independent of pressure for isostatic geometrical families. Deviations only occur near the end of geometrical families.) We find that the affine contribution to the shear modulus for isostatic geometrical families,
\begin{eqnarray}
\label{pressure_dependent}
G^{(1)}_{a}&=& L^{-d}\left(d L^d\right)^{\frac{\alpha-2}{\alpha-1}} P^{\frac{\alpha-2}{\alpha-1}} \times \nonumber \\
& & \sum_{i>j} (\alpha-1)\left(\frac{\sigma_{ij}}{r_{ij}}\right)^{\frac{\alpha-2}{\alpha-1}}\frac{x_{ij}^2y_{ij}^2}{\sigma_{ij}^2r_{ij}^2}\chi_{ij}^{\frac{\alpha-2}{\alpha-1}}  \Theta \left( 1- \frac{r_{ij}}{\sigma_{ij}} \right) \nonumber \\
& & - d P \sum_{i>j} \frac{y_{ij}^4}{r_{ij}^4}\chi_{ij} \Theta \left( 1- \frac{r_{ij}}{\sigma_{ij}} \right),
\end{eqnarray}
has the same form as that for the shear modulus of the first geometrical family, $G^{(1)}$ in Eq.~\ref{family}, i.e. 
\begin{equation}
\label{finale}
G^{(1)}_a = A_a^{(1)}(\alpha) P^{(\alpha-2)/(\alpha-1)} - B_a^{(1)} P,
\end{equation}
where the coefficients $A_a^{(1)}(\alpha)= L^{-d}\left(d L^d\right)^{\frac{\alpha-2}{\alpha-1}} \sum_{i>j} (\alpha-1)\left(\frac{\sigma_{ij}}{r_{ij}}\right)^{\frac{\alpha-2}{\alpha-1}}\frac{x_{ij}^2y_{ij}^2}{\sigma_{ij}^2r_{ij}^2}\chi_{ij}^{\frac{\alpha-2}{\alpha-1}} \Theta \left( 1- \frac{r_{ij}}{\sigma_{ij}} \right)$ and $B_a^{(1)} = d \sum_{i>j} \frac{y_{ij}^4}{r_{ij}^4}\chi_{ij} \Theta \left( 1- \frac{r_{ij}}{\sigma_{ij}} \right)$ are roughly independent of pressure and $B_a^{(1)}$ is independent of the power-law exponent $\alpha$.  For repulsive linear spring interactions, $G^{(1)}_a = A^{(1)}_a(2) - B^{(1)}_a P$ and for repulsive Hertzian spring interactions, $G^{(1)}_a=A^{(1)}_a(5/2) P^{1/3} - B^{(1)}_a P$ as found for $G^{(1)}$ in Sec.~\ref{results:isostatic_families}. 

Since both the $G^{(1)}=G^{(1)}_a - G^{(1)}_n$ and $G^{(1)}_a$ obey the scaling behavior in Eq.~\ref{family}, it is reasonable to assume that the non-affine contribution $G^{(1)}_n$ to the shear modulus for isostatic geometrical families also obeys Eq.~\ref{family}. Thus, the coefficients of the two pressure-dependent terms in Eq.~\ref{family} for $G^{(1)}$ can be decomposed into separate affine and non-affine contributions, i.e. $A^{(1)}_{\alpha} = A^{(1)}_a(\alpha) - A^{(1)}_n(\alpha)$ and $B^{(1)}_{\alpha} = B^{(1)}_a - B^{(1)}_n(\alpha)$.  In Fig.~\ref{fig:Aaffine_Anonaffine_N_all}, we show the distributions of $A^{(1)}_a(5/2)$ and $A^{(1)}_n(5/2)$ for disk packings with repulsive Hertzian spring interactions. The affine and non-affine contributions, $A^{(1)}_a$ and $A^{(1)}_n$, are similar, and thus $A^{(1)}_{\alpha}$ is on average smaller than the magnitudes of the affine and nonaffine contributions separately. In contrast, as shown in Fig.~\ref{fig:Baffine_Bnonaffine_N_all} for disk packings with repulsive Hertzian spring interactions, $B^{(1)}_a > 0$, whereas $\langle B^{(1)}_n(\alpha) \rangle < 0$, and thus the typical $B^{(1)}_{\alpha}$ values are larger than the magnitudes of the affine and nonaffine contributions, $B^{(1)}_a$ and $B^{(1)}_n$, separately.  Also, as shown in the insets to Fig.~\ref{fig:Baffine_Bnonaffine_N_all},  $\langle B^{(1)}_a \rangle$, $\langle -B^{(1)}_n(\alpha) \rangle$, and $\langle B^{(1)}_{\alpha} \rangle$ grow with increasing $N$. As noted in Sec.~\ref{results:isostatic_families}, $G^{(1)}$ begins to deviate from the scaling form in Eq.~\ref{dimensionless2} when the non-affine contribution, $G^{(1)}_n$, does not obey Eq.~\ref{dimensionless2}.

\begin{figure}[h]
\centering
\includegraphics[height=12cm]{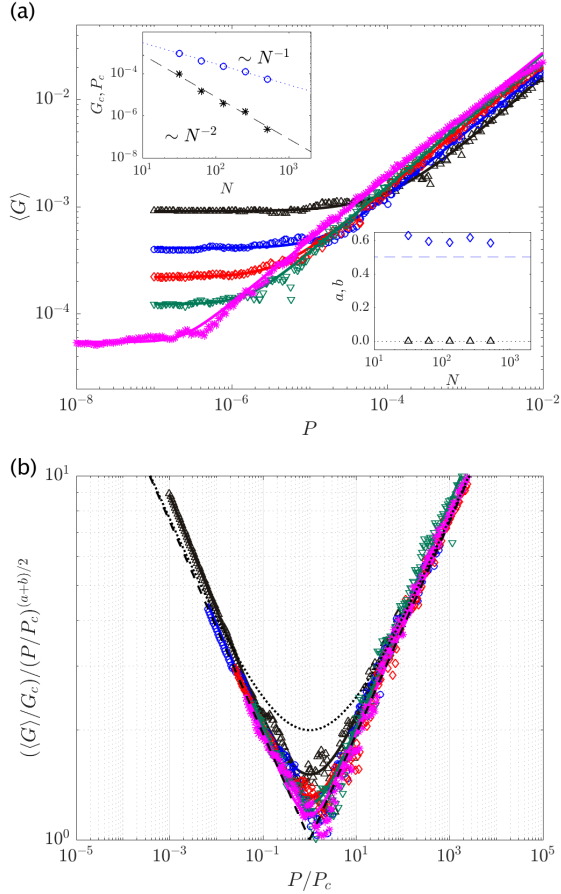}
\caption{(a) Ensemble-averaged shear modulus $\langle G\rangle$ versus $P$ for disk packings with $\alpha=2$ and system sizes $N=32$ (black upward triangles), $64$ (blue circles), $128$ (red diamonds), $256$ (green downward triangles), and $512$ (magenta asterisks). (b) $(\langle G\rangle/G_c) (P/P_c)^{-(a+b)/2}$ versus $P/P_c$ for the data in (a).  The dashed lines have slopes equal to $-1/4$ and $1/4$. The dotted line gives Eq.~\ref{dimensionless3}. The solid lines in (a) and (b) are fits to Eq.~\ref{pnorm} with $p=2$-$5$ for system sizes $N=32$ to $512$. The upper left inset in (a) shows $P_c$ (black asterisks) and $G_c$ (blue circles) versus $N$. The dotted and dashed lines have slopes equal to $-1$ and $-2$, respectively. The lower right inset in (a) gives the exponents, $a$ (black upper triangles) and $b$ (blue diamonds), used in fits to Eq.~\ref{pnorm} versus $N$. The horizontal dotted and dashed lines indicate $a=0$ and $b=0.5$, respectively.}
\label{fig:ensembleG_linear}
\end{figure}

\subsection{Shear Modulus for the Second Geometrical Family}
\label{results:second}

In the previous two sections (Secs.~\ref{results:isostatic_families} and~\ref{results:affine_approximation}), we focused on the pressure-dependent shear modulus $G^{(1)}$ of isostatic geometrical families with $N_c^0$ contacts, prior to the first change in the contact network. In this section, we show preliminary studies of the shear modulus $G^{(2)}$ of the second geometrical family after the packing undergoes a point or jump change in the contact network at $P^*$. (See Figs.~\ref{fig:G_P_second_family_N_32} and~\ref{fig:G_P_second_family_N_32_jump}.) We find that when isostatic geometrical families undergo changes in the contact network during isotropic compression, $\approx 75\%$ undergo point changes to a second geometrical family and $\approx 25\%$ undergo jump changes to a second geometrical family for packings with repulsive linear and Hertzian spring interactions. These fractions do not depend strongly on system size. After point changes, nearly all of the packings in the second geometrical families possess $N_c^0+1$ contacts. For packings (with both $\alpha =2$ and $5/2$ interactions) that undergo jump changes, $\approx 60\%$ of the packings in the second geometrical families possess $N_c^0$ contacts and most of the remaining fraction possess $N_c^0+1$ contacts.  These results also do not depend strongly on system size. 

In Fig.~\ref{fig:G_P_second_family_N_32}, we show the shear modulus $G^{(i)}$ as a function of pressure for a series of disk packings during isotropic compression. At $P^*$, the disk packing (with $\alpha=2$ in (a) and $\alpha=5/2$ in (b)) undergoes a point change and the isostatic geometrical family transitions to a second geometrical family with $N_c^0+1$ contacts. As pointed out in our previous studies~\cite{tuckman}, $G^{(i)}$ is discontinuous across a point change for $\alpha=2$, but it is continuous across a point change for $\alpha >2$. For $\alpha=2$, we find that $G^{(2)}$ for the second geometrical family after a point change obeys the same scaling form in Eq.~\ref{dimensionless2} for isostatic geometrical families, and the characteristic pressure $P_0 \sim N^{-2}$ and shear modulus $G_0 \sim N^{-1}$ for the second geometrical family tend to zero in the large-system limit. (See Fig.~\ref{fig:2nd_families_collapse} (a).) As we found for the first geometrical families in Fig.~\ref{fig:isostatic_collapse} (a), deviations from Eq.~\ref{dimensionless2} can occur at large pressures when $G^{(2)}$ has a significant non-affine contribution and $G^{(2)}_n$ does not obey Eq.~\ref{dimensionless2}.  (See the inset to Fig.~\ref{fig:2nd_families_collapse} (a).)  

The shear moduli for the second geometrical families, $G^{(2)}$, for packings with $\alpha=5/2$ after a point change possess deviations from the scaling form in Eq.~\ref{dimensionless2} both at small pressures near the first point change and at large pressures near the second change in the contact network, as shown in Fig.~\ref{fig:2nd_families_collapse} (b).  For the first geometrical family, there is only one characteristic pressure $P_0$, which determines the crossover pressure that separates the two power-law scaling regimes. For the second geometrical family following a point change with $\alpha>2$, there are two characteristic pressures, $P^*$, indicating the pressure at which the point change from the first to second geometrical family occurs, and $P_0$.  For highly nonlinear interactions with $\alpha>2$, the presence of two characteristic pressures causes $G^{(2)}$ to deviate from Eq.~\ref{dimensionless2}. The inset to Fig.~\ref{fig:2nd_families_collapse} (b) shows that the deviation of $G^{(2)}$ from the scaling form at small pressures is also caused by the non-affine contribution to the shear modulus that does not obey Eq.~\ref{dimensionless2}.  The deviations of $G^{(2)}$ from the scaling form at large pressures following point changes for $\alpha=5/2$ are similar to those found for $G^{(1)}$ near the end of the first geometrical family. 

As shown in Fig.~\ref{fig:G_P_second_family_N_32_jump}, the shear modulus $G^{(i)}$ is discontinuous when the system undergoes a jump change for packings with all $\alpha$.
If an isostatic geometrical family undergoes a jump change to a second geometrical family, $G^{(2)}$ obeys Eq.~\ref{dimensionless2} for packings with $\alpha=2$ and $5/2$ over a wide range of pressure.  (See Fig.~\ref{fig:2nd_families_collapse} (c).) Again, there can be deviations in $G^{(2)}$ from the scaling form at large pressures when the non-affine contribution does not obey Eq.~\ref{dimensionless2}. In future studies, we will investigate the general form of the shear modulus $G^{(i)}(P)$ for the third, fourth, and higher-order geometrical families at elevated pressures.   

\begin{figure}[h]
\centering
\includegraphics[height=12cm]{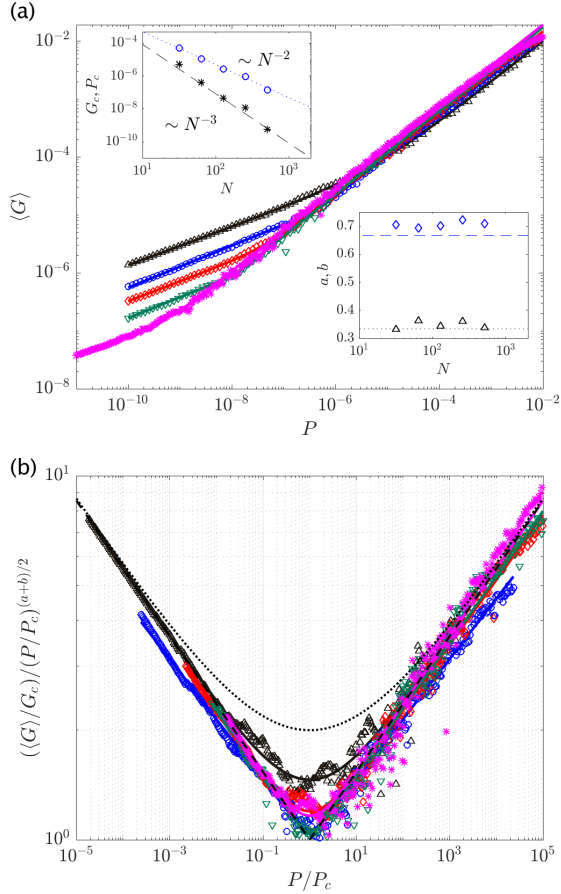}
\caption{(a) Ensemble-averaged shear modulus $\langle G\rangle$ versus $P$ for disk packings with $\alpha=2.5$ and system sizes $N=32$ (black upward triangles), $64$ (blue circles), $128$ (red diamonds), $256$ (green downward triangles), and $512$ (magenta asterisks). (b) $(\langle G\rangle/G_c) (P/P_c)^{-(a+b)/2}$ versus $P/P_c$ for the same data in (a). The dashed lines have slopes equal to $-1/6$ and $1/6$. The dotted line gives Eq.~\ref{dimensionless3}. The solid lines in (a) and (b) are fits to Eq.~\ref{pnorm} with $p=2$-$15$ for system sizes $N=32$ to $512$. The upper left inset in (a) shows $P_c$ (black asterisks) and $G_c$ (blue circles) versus $N$. The dotted and dashed lines have slopes equal to $-2$ and $-3$, respectively. The lower right inset in (a) shows the exponents, $a$ (black upper triangles) and $b$ (blue diamonds), used in fits to Eq.~\ref{pnorm} versus $N$. The horizontal dotted and dashed lines indicate $a=1/3$ and $b=2/3$, respectively.}

\label{fig:ensembleG_Hertzian}
\end{figure}

\subsection{Ensemble-averaged Shear Modulus}
\label{results:rearrangement_contribution}

In this section, we investigate the pressure dependence of the ensemble-averaged shear modulus $\langle G \rangle$, which is often studied to mimic the large-system limit.  As shown in the previous section, jump changes in the contact network give rise to discontinuities in the shear modulus for packings with all $\alpha$. In contrast, the shear modulus is continuous across point changes for $\alpha>2$, but it is discontinuous for $\alpha=2$.  The shear modulus for a single initial condition $\lambda$ at $P=0$ undergoing isotropic compression can be written as $G^{\lambda} = G^{\lambda}_f + G^{\lambda}_r$, where $G^{\lambda}_f$ describes the shear modulus along continuous geometrical families and $G^{\lambda}_r$ includes discontinuities in the shear modulus from point and jump changes. (See Fig.~\ref{fig:G_P_entire_single_all_N_32}.) $G_r$ for $\alpha=2$ includes discontinuities in the shear modulus from both point and jump changes, whereas $G_r$ includes changes in the shear modulus from jump changes only for $\alpha >2$.  The ensemble-averaged shear modulus, $\langle G \rangle$, is obtained by averaging over initial conditions.

In Fig.~\ref{fig:decouplingDemo}, we show $\langle G\rangle$, $|\langle G_f \rangle |$, and $\langle G_r\rangle$ for $N=128$ disk packings with $\alpha =2$ and $5/2$.  At small pressures, $\langle G \rangle \sim \langle G_f\rangle$ since changes in the contact network are rare. In the $P \rightarrow 0$ limit, $\langle G \rangle$ is a constant for packings with $\alpha=2$ and $\langle G\rangle \sim P^{1/3}$ for packings with $\alpha=5/2$, consistent with the results in Sec.~\ref{results:isostatic_families}. For packings with $\alpha =2$ and $5/2$, as the pressure increases, $\langle G_f \rangle$ decreases toward zero and at a characteristic pressure, $\langle G\rangle \approx \langle G_r\rangle$.  As the pressure continues to increase, $\langle G_f \rangle <0$ (since the $-BP$ term dominates the geometrical family contribution at large pressures), which causes the cusp in $|\langle G_f\rangle |$ in Fig.~\ref{fig:decouplingDemo}. At large pressures, both $\langle G_f \rangle$ and $\langle G_r \rangle$ contribute to $\langle G\rangle$, and $\langle G \rangle < \langle G_r\rangle$. 

In contrast to the scaling behavior found for the shear modulus $G^{(1)}$ of isostatic geometrical families, the pressure dependence of the {\it ensemble-averaged} shear modulus $\langle G\rangle$ is not simply the sum (or difference) of two power-laws in pressure~\cite{vanderwerf}, $\langle G\rangle \sim A P^a + B P^b$ with exponents $a$ and $b$.  (See  Figs.~\ref{fig:ensembleG_linear} and~\ref{fig:ensembleG_Hertzian}.) To illustrate this, we consider the dimensionless, symmetric form for $\langle G\rangle$: 
\begin{equation}
\label{dimensionless3}
\frac{\langle G\rangle}{G_c} \left( \frac{P}{P_c} \right)^{-(a+b)/2}  =  \left( \frac{P}{P_c} \right)^{(a-b)/2} 
+ \left( \frac{P}{P_c} \right)^{-(a-b)/2},
\end{equation}
where the scaling is dominated by $P^{(a-b)/2}$ for $P<P_c$ and by $P^{-(a-b)/2}$ for $P > P_c$.  In Figs.~\ref{fig:ensembleG_linear} (b) and~\ref{fig:ensembleG_Hertzian} (b), we plot Eq.~\ref{dimensionless3} as dashed lines for packings with $\alpha=2$ and $5/2$ and compare it to the simulation results for $\frac{\langle G\rangle}{G_c(\alpha)} \left( \frac{P}{P_c(\alpha)} \right)^{-(a+b)/2}$ for $\alpha \sim (\alpha-2)/(\alpha-1)$ and $b \sim (\alpha-3/2)/(\alpha-1)$.  For packings with both $\alpha=2$ and $5/2$, the simulation data transitions between the two limiting power-law behaviors $(P/P_c)^{(a-b)/2}$ and $(P/P_c)^{-(a-b)/2}$ much more abruptly than the sum of the two power-laws, $(P/P_c)^{(a-b)/2} + (P/P_c)^{-(a-b)/2}$.
To capture this feature in the simulation data, we fit the simulation data to the $p$-norm of the right-hand side of Eq.~\ref{dimensionless3}, i.e.
\begin{equation}
\label{pnorm}
\frac{\langle G\rangle}{G_c} \left( \frac{P}{P_c} \right)^{-\frac{a+b}{2}} = \left[\left( \frac{P}{P_c} \right)^{\frac{p(a-b)}{2}} + \left( \frac{P}{P_c} \right)^{-\frac{p(a-b)}{2}} \right]^{\frac{1}{p}},
\end{equation}
with $p \sim 2$-$5$ ($\sim 2$-$15$) for packings with $N=32$ to $512$ and $\alpha=2$ ($\alpha=5/2$). The $p$-norm generates polynomials with powers between $(a-b)/2$ and $-(a-b)/2$ to capture the kink-like feature in the simulation data. In the insets to Figs.~\ref{fig:ensembleG_linear} (a) and~\ref{fig:ensembleG_Hertzian} (a), we show that the exponent $a\sim (\alpha-2)/(\alpha-1)$ controls the low-pressure behavior of $\langle G\rangle$. However, using best fits to Eq.~\ref{pnorm}, the exponent $b \gtrsim (\alpha-3/2)/(\alpha-1)$, which controls the large pressure behavior. Further studies of $\langle G\rangle$ for larger system sizes are necessary to more accurately determine the exponent $b$.   By fitting Eq.~\ref{pnorm}, we also find that  
$P_c \sim N^{-2(\alpha-1)}$ and $G_c \sim N^{-2(\alpha-3/2)}$. (See the insets to  Figs.~\ref{fig:ensembleG_linear} (a) and~\ref{fig:ensembleG_Hertzian} (a).)

Using Eq.~\ref{pnorm}, we can solve for $\langle G\rangle$ as a function of $P$.  We display the system-size dependence of $\langle G(P) \rangle$ for disk packings with repulsive linear and Hertzian spring interactions in Figs.~\ref{fig:ensembleG_linear} (a) and~\ref{fig:ensembleG_Hertzian} (a), respectively. As we found in Figs.~\ref{fig:ensembleG_linear} (b) and~\ref{fig:ensembleG_Hertzian} (b), Eq.~\ref{pnorm} provides an excellent fit for $\langle G(P) \rangle$ for all system sizes. At sufficiently low pressures, we find that 
$\langle G \rangle \sim P^a$, where $a\sim (\alpha-2)/(\alpha-1)$, whereas $\langle G \rangle \sim P^{b}$ at high pressures, where $b \gtrsim (\alpha-3/2)/(\alpha-1)$. For $\alpha=2$, the scaling exponents in the low- and high-pressure limits are $a \sim 0$ and $b \sim 0.60$, and for $\alpha=5/2$, the scaling exponents in the low- and high-pressure limits are $a \sim 0.36$ and $b\sim 0.70$~\cite{ellenbroek}.  In the large-$\alpha$ limit, we predict that the scaling exponents in the low- and high-pressure limits will both approach $1$.  We find similar behavior for sphere packings in 3D for $\alpha=2$ and $5/2$. 

\section{Conclusions and Future Directions} 
\label{conclusions}
The mechanical response of jammed packings of purely repulsive spherical particles to isotropic compression is complex~\cite{roux,wyart2}. For example, several studies have shown that effective medium theory, which assumes an affine response to applied deformation, does not accurately predict the behavior of the shear modulus as a function of pressure~\cite{makse,makse2}. In addition, simulations of the "soft particle" model~\cite{jamming}, which assumes purely repulsive, finite-ranged interactions between spherical particles that scale as a power-law in their overlap with exponent $\alpha$, have suggested that the ensemble-averaged shear modulus scales with pressure as $\langle G \rangle \sim P^{(\alpha -3/2)/(\alpha-1)}$. However, the origin of the scaling exponent $(\alpha-3/2)/(\alpha-1)$ for the ensemble-averaged shear modulus is not well-understood.    

In a recent study, we showed that there are two important contributions to the shear modulus in jammed packings of spherical particles undergoing isotropic compression~\cite{vanderwerf}: continuous variations in the shear modulus from geometrical families, for which the interparticle contact network does not change, and discontinuous jumps in the shear modulus from changes in the contact network.  In the present work, we show explicitly for $\alpha=2$, $5/2$, and $3$ that the form of the shear modulus versus pressure for the first, isostatic geometrical family can be inferred from the affine shear response, i.e. $G^{(1)} = A^{(1)}_{\alpha} P^{(\alpha -2)/(\alpha-1)} - B^{(1)}_{\alpha} P$, but the values of the coefficients $A^{(1)}_{\alpha}$ and $B^{(1)}_{\alpha}$, are strongly affected by the non-affine contribution.  

For each initial configuration at $P \sim 0$ that we isostropically compress, we can decompose the shear modulus $G= G_f + G_r$ into contributions from geometrical families ($G_f$) and from discontinuities arising from point and jump changes in the contact network ($G_r$). We show that the ensemble-averaged shear modulus $\langle G \rangle \sim \langle G_f \rangle$ at low pressures since changes in the contact network are rare.  At larger pressures, the geometrical family contribution is dominated by the $-B P$ term (or other higher-order negative terms), $\langle G_f \rangle < 0$, and $\langle G \rangle < \langle G_r \rangle$.  We find that both $\langle G_f \rangle$ and $\langle G_r \rangle$
are important for determining $\langle G\rangle$ at finite pressure in the large-system limit.  Further, we show that the pressure dependence of $\langle G\rangle$ is not simply a sum of two power-laws over the full range of pressure, but $\langle G \rangle \sim P^{(\alpha -2)/(\alpha-1)}$ in the $P \rightarrow 0$ limit, $\langle G \rangle \sim P^b$ at large pressures, where $b \gtrsim (\alpha -3/2)/(\alpha-1)$, and the characteristic pressure that separates these scaling regimes, $P_c \sim N^{-1/[2(\alpha-1)]}$, tends to zero in the large-system limit. 

This work suggests several new areas for future research. First, we investigated the pressure-dependence of the shear modulus for the first, isostatic geometrical family and provided preliminary results for the shear modulus of the second geometrical family with $N_c^0+1$ contacts.  However, we do not yet know the pressure dependence of the shear modulus for higher-order geometrical families that occur at higher pressures.  The answer to this question is crucial for developing a theoretical description for the mechanical response of jammed packings undergoing isotropic compression. Second, numerical simulations suggest that the ensemble-averaged shear modulus for packings of frictional spherical particles has similar pressure dependence as packings of frictionless spherical particles, scaling roughly as $\langle G\rangle \sim P^{(\alpha-3/2)/(\alpha-1)}$ at large pressures~\cite{somfai2007critical}. However, the separate contributions to the shear modulus from geometrical families and changes in the contact network have not yet been studied for packings of {\it frictional} spherical particles. For example, does the shear modulus have the same pressure dependence across point and jump changes~\cite{tuckman} for packings of frictional particles? Third, several computational studies have shown that $\langle G\rangle \sim P^{\kappa}$ at large pressures for non-spherical particles (such as packings of ellipses and circulo-lines~\cite{mailman,schreck}) with purely repulsive linear spring interactions ($\alpha=2$), where $0.5 < \kappa <1$. These results suggest that the scaling exponent for $\langle G(P) \rangle$ at large pressures depends on both the particle shape~\cite{wyart} (e.g. aspect ratio ${\cal A}$) and $\alpha$. It will be interesting to calculate $\langle G(P) \rangle$ for packings of elongated shapes as a function of aspect ratio to determine $\kappa({\cal A},\alpha)$ to understand how the rotational degrees of freedom affect the mechanical response. Further, for packings of non-spherical particles undergoing isotropic compression, there have not been detailed studies of the separate contributions to $\langle G\rangle$ from geometrical families and from changes in the contact network.  A first step would be to calculate the affine shear response for packings of non-spherical particles since for spherical particles the affine response provides insight into the geometrical family contribution to the shear modulus.    

\begin{figure}[h]
\centering
\includegraphics[height=12cm]{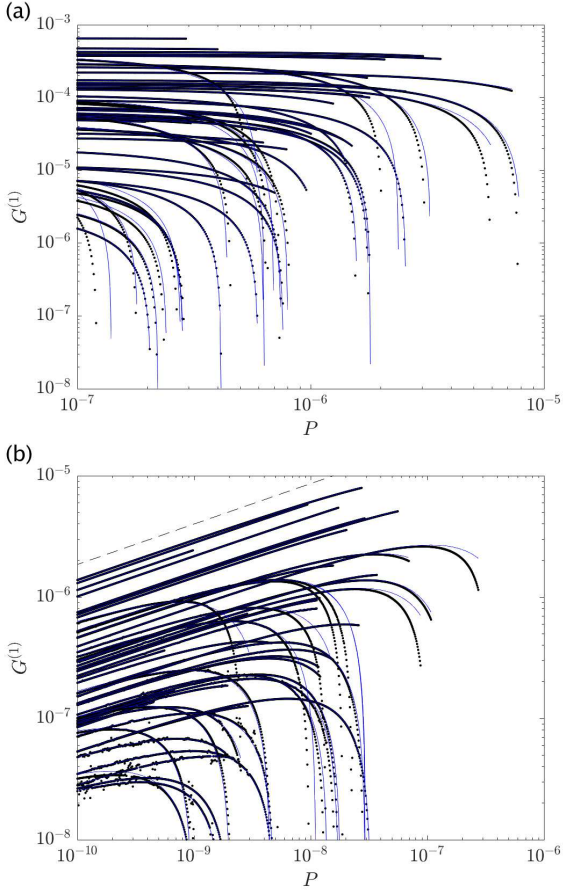}
\caption{Shear modulus $G^{(1)}$ within isostatic geometrical families versus pressure $P$ for individual $N=64$ sphere packings with (a) $\alpha=2$ and (b) $5/2$ repulsive interactions. The solid blue lines are fits to Eq.~\ref{family}. The dashed line in (b) has slope equal to $1/3$.}
\label{fig:G_P_3D_N_64}
\end{figure}

\begin{acknowledgments}
We acknowledge support from the Army Research Laboratory under Grant No. W911NF-17-1-0164 (P. W., N. O., and C. O.), NSF Grant No. DBI-1755494 (P. T.), and China Scholarship Council Grant No. 201906340202 (S. Z.). This work was also supported by the High Performance Computing facilities operated by Yale's Center for Research Computing and computing resources provided by the Army Research Laboratory Defense University Research Instrumentation Program Grant No. W911NF-18-1-0252. 
\end{acknowledgments}

\appendix
\section*{Appendix A}
\label{app:A}
\begin{figure}[h]
\centering
\includegraphics[height=6cm]{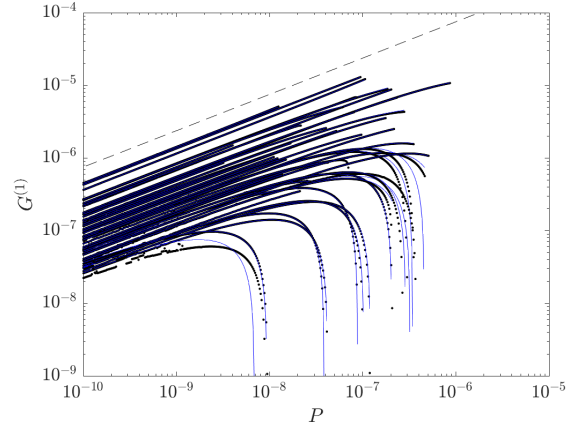}
\caption{Shear modulus $G^{(1)}$ within isostatic geometrical families versus pressure $P$ for individual $N=32$ disk packings with repulsive interactions with power-law exponent $\alpha=3$ in Eq.~\ref{potential}. The dashed line has slope equal to $1/2$. The solid blue lines are fits to Eq.~\ref{family} for each of the $50$ geometrical families.}
\label{fig:G_P_N_32_alpha_3}
\end{figure}

In this Appendix, we provide the results for the shear modulus for isostatic geometrical families for 2D packings of purely repulsive disks with $\alpha=3$ and 3D packings of purely repulsive spheres with $\alpha=2$ and $5/2$.  In Fig.~\ref{fig:G_P_3D_N_64}, we show that $G^{(1)}$ for sphere packings with $\alpha=2$ and $5/2$ also obeys Eq.~\ref{family}. $G^{(1)}$ for $\alpha=2$ is constant and $G^{(1)}$ for $\alpha=5/2$ scales as $P^{1/3}$ in the $P\rightarrow 0$ limit and 
$G^{(1)}$ begins decreasing at larger pressures. In Fig.~\ref{fig:G_P_N_32_alpha_3}, we show that $G^{(1)}$ for disk packings with $\alpha=3$ obeys Eq.~\ref{family}, scaling as $P^{1/2}$ in the $P\rightarrow 0$ limit and then decreasing at larger pressures.  (Note that for some of the geometrical families for $\alpha=3$, a change in the contact network occurs before the $-BP$ term begins contributing significantly to $G^{(1)}$.) 

\section*{Appendix B}
\label{app:B}

In this Appendix, we derive expressions for the affine and non-affine contributions~\cite{maloney,zaccone2011} to the shear modulus $G^{(i)}$ of isostatic geometrical families. (We consider 2D systems here, but a similar derivation holds for 3D systems.) When we apply an affine simple shear deformation, the particle positions are transformed to $(x^{a}_{i},y^{a}_{i})=(x^{0}_{i}+\gamma y^{0}_{i}, y^{0}_{i})$ consistent with Lees-Edwards periodic boundary conditions, where $(x^{0}_{i},y^{0}_{i})$ are the particle positions in the undeformed, reference jammed packing. After each simple shear strain increment $\gamma$, we minimize the total potential energy $U$ at constant packing fraction. Thus, after relaxation, the positions of the particles can be written as the sum of an affine term plus a nonaffine term caused by energy minimization:
\begin{equation}
	\left(x^{'}_{i},y^{'}_{i}\right)=\left(x^{0}_{i}+\gamma y^{0}_{i} + x^{n}_{i}, y^{0}_{i}+y^{n}_{i}\right).
\end{equation}
For each reference packing, we can write the total potential energy as a function of the shear strain and non-affine particle positions, $r^{n}_{i\beta}$, where $i=1,2,\ldots,N$ indicates the particle index and $\beta=x,y$ indicates the Cartesian component of ${\vec r}$. We assume that the disk packing is at an energy minimum after each shear strain increment and the total force on each particle remains zero. Thus, 
\begin{equation}
\label{forcea}
f_{i \beta} = -\left( \frac{\partial U}{\partial r_{i\beta}} \right)_{\gamma}=-\left( \frac{\partial U}{\partial r^{n}_{i \beta}}\right)_{\gamma}=0,
\end{equation}
where $(.)_{\gamma}$ indicates that the derivatives are evaluated at a fixed shear strain $\gamma$. We can then take the derivative of Eq.~\ref{forcea} with respect to $\gamma$,
\begin{eqnarray}
\label{balance1}
-\frac{df_{i \beta}}{d\gamma} &= &\left(\frac{\partial^{2}U}{\partial r^{n}_{i \beta}\partial \gamma} \right) + \left(\frac{\partial^{2}U}{\partial r^{n}_{i \beta}\partial r^{n}_{j \beta}}\right) \frac{dr^{n}_{j \beta}}{d\gamma}\\
\label{balance2}
&=&\left(\frac{\partial^{2}U}{\partial r^{n}_{i \beta}\partial \gamma}\right) + \left(\frac{\partial^{2}U}{\partial r_{i \beta}\partial r_{j \beta}}\right) \frac{dr^{n}_{j \beta}}{d\gamma}=0.
\end{eqnarray}
The shear stress $\Sigma$ is related to the total derivative of the potential energy with respect to $\gamma$ (Eq.\ref{stress2}):
\begin{equation}
\label{stress3}
\Sigma L^d = \frac{dU}{d\gamma}= \frac{\partial U}{\partial r^{n}_{i \beta}} \frac{dr^{n}_{i \beta}}{d\gamma} + \frac{\partial U}{\partial \gamma} = \frac{\partial U}{\partial \gamma}.  
\end{equation}
Note that for a given reference configuration at fixed strain $\gamma$, taking derivatives with respect to $r_{i \beta}$ is equivalent to taking derivatives with respect to $r^{n}_{i \beta}$. 

Using Eq.~\ref{balance2}, we can solve for the derivative, $dr^n_{i \beta}/d\gamma$:
\begin{equation}
\frac{dr^{n}_{i \beta}}{d\gamma} = -M^{-1}_{ij}\Xi_{j \beta},
\end{equation}
where
\begin{equation}
\Xi_{i \beta}=\frac{\partial^{2}U}{\partial r^{n}_{i \beta}\partial \gamma}
\end{equation}
and the Hessian matrix $M_{ij}$ is defined by the second derivatives of the total potential energy $U$ with respect to the particle coordinates,
\begin{equation}
M_{ij} = \frac{\partial^{2}U}{\partial r_{i \beta}\partial r_{j \beta}}.
\end{equation}

Using Eq.~\ref{stress3}, we can calculate the shear modulus $G L^d = d\Sigma/d\gamma$, 
\begin{eqnarray}
\label{shear_modulus}
G L^d &= &\frac{d}{d\gamma} \left(\frac{\partial U}{\partial \gamma} + \frac{\partial U}{\partial r^{n}_{i \beta}}\frac{dr^{n}_{ i \beta}}{d\gamma}\right)\\
& = &\frac{\partial^{2}U}{\partial\gamma^{2}} + \frac{\partial^{2}U}{\partial r^{n}_{i \beta}\partial\gamma}\frac{dr^{n}_{i \beta}}{d\gamma}\\
& = & \frac{\partial^{2}U}{\partial\gamma^{2}} - \Xi_{i \beta}M^{-1}_{ij}\Xi_{j \beta}.
\end{eqnarray}
Thus, we find that the shear modulus $G = G^{a}-G^{n}$, where $G^a = L^{-d} \partial^2 U/\partial \gamma^2$ is the affine contribution and $G^n = L^{-d} \Xi_{i \beta}M^{-1}_{ij}\Xi_{j \beta}$ is the non-affine contribution.

\begin{figure}[h]
\centering
\includegraphics[height=12cm]{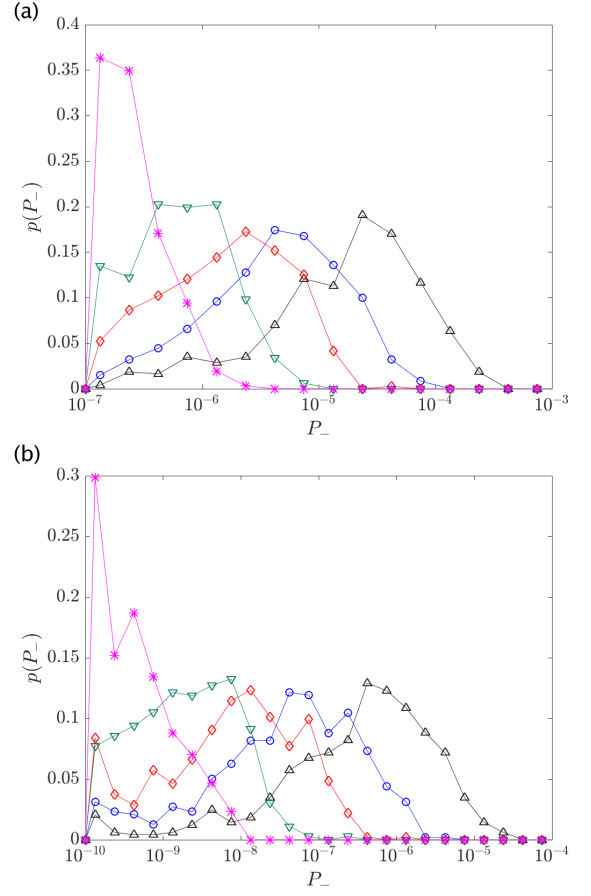}
\caption{The probability distribution $p(P_{-})$ of the pressure $P_{-}$ at which the shear modulus for the isostatic geometrical family first becomes negative $G^{(1)} <0$ for disk packings with (a) $\alpha=2$ and (b) $5/2$ and system sizes $N=32$ (black upward triangles), $64$ (blue circles), $128$ (red diamonds), $256$ (green downward triangles), and $512$ (magenta asterisks). }
\label{fig:P_dist_1st_neg_G_N_all}
\end{figure}

\section*{Appendix C}
\label{app:C}

\begin{figure}[h]
\centering
\includegraphics[height=12cm]{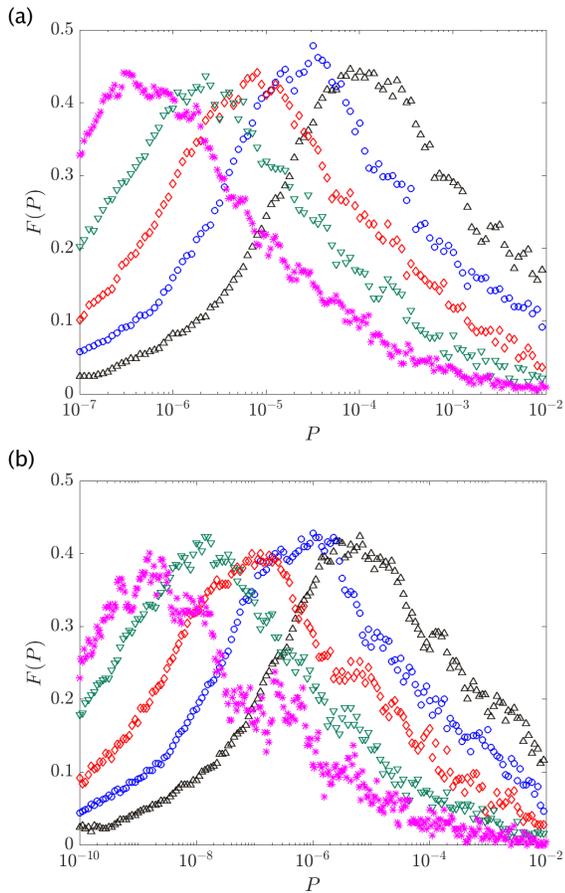}
\caption{Fraction of packings $F(P)$ at each pressure that possess a negative shear modulus for disk packings with (a) $\alpha=2$ and (b) $5/2$ and system sizes $N=32$ (black upward triangles), $64$ (blue circles), $128$ (red diamonds), $256$ (green downward triangles), and $512$ (magenta asterisks).}
\label{fig:find_neg_G_all_states_N_all}
\end{figure}

In this Appendix, we quantify the frequency with which disk packings generated using the strain-controlled energy minimization method possess negative shear moduli. (We have verified that all packings generated via isotropic compression, even those with negative shear moduli, possess positive bulk moduli.) The shear modulus for the first geometrical family obeys Eq.~\ref{family}; $G^{(1)} >0$ in the $P\rightarrow 0$ limit, but it decreases strongly with increasing pressure. Thus, the shear modulus can become negative if a point or jump change in the contact network does not occur abruptly after the start of the first geometrical family. In Fig.~\ref{fig:P_dist_1st_neg_G_N_all}, we show the distribution $p(P_-)$ of the pressure $P_-$ at which the isostatic geometrical family first becomes negative. We find that $\langle P_- \rangle \sim P_0$ and thus $\langle P_-\rangle$ tends to zero in the large-system limit. $\langle P_- \rangle \sim N^{-2}$ and $\sim N^{-3}$ for packings with $\alpha =2$ and $5/2$, respectively.  After a jump change and after a point change for $\alpha =2$, the shear modulus for the second geometrical family $G^{(2)}$ jumps discontinuously to either a positive or negative value, depending on the value of $G^{(1)}$ at the end of the first geometrical family and the magnitude and sign of the discontinuous jump in the shear modulus. As the pressure increases, the upward jumps in the shear modulus become larger than the continuous decreases in the shear modulus along geometrical families, and thus the shear modulus remains positive.  In Fig.~\ref{fig:find_neg_G_all_states_N_all}, we show the fraction of disk packings $F(P)$ at each pressure with a negative shear modulus.  The maximum fraction of packings with negative shear moduli is $\approx 0.4$ and occurs at $P_{\rm max}/P_c \approx 1$. Thus, $P_{\rm max} \sim N^{-2}$ and $\sim N^{-3}$ for $\alpha=2$ and $5/2$, respectively.      

\bibliography{rsc.bib}
\bibliographystyle{rsc}

\end{document}